\documentclass[10pt]{article}
\pdfoutput=1
\usepackage{jheparxiv}
\usepackage[latin1]{inputenc}
\usepackage{latexsym,diagbox,stackrel}
\usepackage{relsize}
\usepackage{amsmath}
\usepackage{mathrsfs}
\usepackage{arydshln,leftidx,mathtools}
\usepackage{bbold}
\usepackage{pdflscape}
\usepackage{amssymb}
\usepackage{multirow}
\usepackage{enumerate}
\usepackage[usenames,dvipsnames,svgnames,table]{xcolor}
\usepackage{cleveref}
\usepackage{tikz}
\usepackage{slashed}
\usepackage[debug]{mciteplus}

\def\d{\text{d}}

\newcommand{\C}{\mathbb{C}}
\newcommand{\CP}{\mathbb{CP}}

\newcommand{\cI}{\mathcal{I}}

\newcommand{\Z}{\mathbb{Z}}

\newcommand{\vol}{\mathrm{vol}\,}

\newcommand{\SL}{\mathrm{SL}}

\newcommand{\tr}{\mathrm{tr}}
\newcommand{\sgn}{\mathrm{sgn}}

\newcommand{\pf}{\text{Pf} \,}

\newcommand{\RO}{\mathrm{RO}}

\subheader{\hfill \begin{tabular}{r} \texttt{QMUL-PH-17-29} \end{tabular}}
 
\title{\Large Gluons and gravitons at one loop from ambitwistor strings}
\author[a,c]{Yvonne Geyer}
\author[b,c]{\& Ricardo Monteiro}
        
\affiliation[a]{School of Natural Sciences, Institute for Advanced Study \\
        Einstein Drive, Princeton, NJ 08540, USA}

\affiliation[b]{Centre for Research in String Theory \\
        Queen Mary University of London, E1 4NS, United Kingdom}

\affiliation[c]{Kavli Institute for Theoretical Physics \\
        University of California Santa Barbara, CA 93106-4030, USA
}
        
\emailAdd{yvonnegeyer@ias.edu}
\emailAdd{ricardo.monteiro@qmul.ac.uk}

\abstract{We present new and explicit formulae for the one-loop integrands of scattering amplitudes in non-supersymmetric gauge theory and gravity, valid for any number of particles. The results exhibit the colour-kinematics duality in gauge theory and the double-copy relation to gravity, in a form that was recently observed in supersymmetric theories. The new formulae are expressed in a particular representation of the loop integrand, with only one quadratic propagator, which arises naturally from the framework of the loop-level scattering equations. The starting point in our work are the expressions based on the scattering equations that were recently derived from ambitwistor string theory. We turn these expressions into explicit formulae depending only on the loop momentum, the external momenta and the external polarisations. These formulae are valid in any number of spacetime dimensions for pure Yang-Mills theory (gluon) and its natural double copy, NS-NS gravity (graviton, dilaton, B-field), and we also present formulae in four spacetime dimensions for pure gravity (graviton). We perform several tests of our results, such as checking gauge invariance and directly matching our four-particle formulae to previously known expressions. While these tests would be  elaborate in a Feynman-type representation of the loop integrand, they become straightforward in the representation we use.
}

\begin{document}

\maketitle


\section{Introduction}
Recent progress in the study of perturbative scattering amplitudes has often relied on supersymmetry. While the study of supersymmetric theories has both theoretical and practical motivations, it is important to know to what extent recent findings actually rely on supersymmetry, not least in view of the relevance to present day phenomenology. In this work, we study non-supersymmetric gauge theory and gravity by employing two remarkable developments in the study of amplitudes: (i) the formalism of the scattering equations and ambitwistor strings, and (ii) the double-copy relation between gauge theory and gravity. These closely interconnected structures are easier to study at loop level when working with supersymmetric theories. They are, however, also expected to hold in the absence of supersymmetry. This is indeed what we find.

Our main outcome is a set of explicit expressions for the $n$-particle one-loop integrands in non-supersymmetric gauge theory and gravity. The formulae are a stepping stone in the application of ambitwistor strings to practical calculations of scattering amplitudes, including for theories of phenomenological interest. Moreover, our results have consequences for the understanding of gravity as the `square' of non-abelian gauge theories.

In its basic structure, the double-copy relation between Yang-Mills theory and gravity remains intriguing, especially at loop level where it is only a conjecture. The most fruitful tool to understand it has been the Bern-Carrasco-Johansson (BCJ) duality between colour and kinematics \cite{Bern:2008qj, Bern:2010ue}. The {\it colour-kinematics duality} ascertains a hidden symmetry of the gauge-theory S-matrix when expressed as a sum over cubic Feynman-like diagrams. Specifically, it states that the kinematic numerators associated to the diagrams can be chosen to satisfy a Jacobi relation whenever the corresponding colour factors satisfy that Jacobi identity. This property ensures that gravity amplitudes can be obtained from gauge theory amplitudes by simply substituting the colour-factor by another copy of the kinematic factor, a process known as {\it double copy}. At tree level, the BCJ double copy is equivalent to the Kawai-Lewellen-Tye (KLT) relations \cite{Kawai:1985xq} arising from string theory. String theory provides a beautiful interpretation of the double copy as relating closed string amplitudes to open string amplitudes; where gravitons, as massless vibrations of closed strings, arise from joining the endpoints of a pair of open strings. In fact, the tree-level BCJ structure has been proven by the same worldsheet monodromy properties of string theory that underlie the KLT relations \cite{BjerrumBohr:2009rd,*Stieberger:2009hq}; and it has also been proven using modern field theory techniques \cite{Bern:2010yg,*Feng:2010my}. There are numerous constructions of tree-level BCJ representations for gauge theory and gravity, e.g.~\cite{Tye:2010dd,Mafra:2011kj,Monteiro:2011pc,*BjerrumBohr:2012mg,Oxburgh:2012zr,Broedel:2012rc,Johansson:2014zca,*Johansson:2015oia,Chiodaroli:2014xia,*Chiodaroli:2015rdg,*Chiodaroli:2017ehv,Anastasiou:2017nsz}. The double-copy structure also turns out 
to generalise to a variety of theories -- and relations between them -- ranging from open string theory to effective scalar field theories such as the non-linear sigma model and the Galileon \cite{Mafra:2011nv,*Broedel:2013tta,Carrasco:2016ldy, Cachazo:2014xea,Cheung:2017ems}. Moreover, it applies beyond flat-space scattering amplitudes, to perturbative (and in certain cases exact) solutions to the classical equations of motion, e.g.~\cite{Monteiro:2014cda,Goldberger:2016iau,Luna:2016hge,*Luna:2017dtq,Anastasiou:2014qba,*Cardoso:2016amd}, and to scattering on curved backgrounds \cite{Adamo:2017nia,*Adamo:2017sze}.

At loop level, the straightforward extension of the colour-kinematics duality to the loop integrand is conjectural \cite{Bern:2010ue}. There are many non-trivial examples of supersymmetric amplitudes for which a BCJ representation exists, to cite a few \cite{Bern:2010ue,Carrasco:2011mn,Bern:2012uf, Bern:2012cd, *Bern:2013qca, *Bern:2013uka, *Bern:2014sna,Mafra:2015mja,Johansson:2017bfl}, and there are also examples of form factors \cite{Boels:2012ew,Yang:2016ear}. In fact, at one loop there is strong evidence that the colour-kinematics duality holds in general, and that the underlying structure is similar to that at tree level (irrespective of supersymmetry) \cite{Boels:2011tp, *Boels:2011mn, Du:2012mt, Mafra:2012kh, Boels:2013bi, Bjerrum-Bohr:2013iza,  Bern:2013yya, Mafra:2014gja, He:2015wgf, Primo:2016omk, Jurado:2017xut}, including the string theory monodromy story \cite{Tourkine:2016bak,*Hohenegger:2017kqy,*Ochirov:2017jby}.

However, there are also obstacles to the colour-kinematics duality at generic loop level. In particular, it may not be possible to find a colour-kinematics satisfying representation if one restricts the numerators to be local and to satisfy reasonable power counting properties in the loop momenta. The most notable example of this difficulty is the five-loop four-point amplitude in maximal super-Yang-Mills theory. The motivation is the construction of its double copy in maximal supergravity, whose ultraviolet properties -- a long standing problem -- may offer hints on the all-loop ultraviolet behaviour of the theory. Recent work \cite{Bern:2017yxu} has explored the double-copy structure in the absence of numerators satisfying the Jacobi relations, and this has allowed for the construction of the full loop integrand for this amplitude \cite{Bern:2017ucb}, although more work is required to extract the ultraviolet behaviour. There are previous examples of getting around this type of BCJ obstacle by relaxing in some way the Jacobi relations \cite{Yuan:2012rg,Bern:2015ooa} or the loop power counting \cite{Mogull:2015adi}. Investigations from the field theory limit of string theory have also identified obstructions to the naive expectations of the BCJ structure \cite{Mafra:2014gja, Berg:2016fui}.

Given these obstacles and potential solutions, it is important to revisit the workings of the colour-kinematics duality using a first-principles approach. Such an approach is provided by the other main development that we explore in this work, namely the Cachazo-He-Yuan (CHY) formalism of the scattering equations \cite{Cachazo:2013gna,Cachazo:2013hca,Cachazo:2013iea} and its worldsheet interpretation in terms of ambitwistor strings, discovered by Mason and Skinner \cite{Mason:2013sva,Geyer:2014fka,Adamo:2014wea,Ohmori:2015sha,Casali:2015vta,Adamo:2017zkm,Adamo:2017nia,*Adamo:2017sze}. The latter are quantum field theories formulated in a string-like manner, and they inherit with some modifications the structures of ordinary string theory that so naturally express the double copy; refs.~\cite{Siegel:2015axg, *Casali:2016atr, *Casali:2017zkz,*Casali:2017mss, *Yu:2017bpw, Azevedo:2017lkz, *Azevedo:2017yjy} study the precise connection to string theory. Crucially, ambitwistor strings provide a loop-level framework which is much easier to work with than the genus expansion of ordinary string theory. While some ambitwistor string models admit a genus expansion \cite{Adamo:2013tsa,Casali:2014hfa,Adamo:2015hoa}, there is a more general formalism of an expansion in the number of nodes of a Riemann sphere, with the nodes (pairs of identified points) representing the loop momenta \cite{Geyer:2015bja,Geyer:2015jch,Geyer:2016wjx,Roehrig:2017gbt}. This expansion can in principle be derived from the genus expansion when the latter exists, and in fact it was discovered in this way at one loop \cite{Geyer:2015bja}. For an alternative approach to the loop-level scattering equations, based on a (hyper)elliptic curve, see \cite{Gomez:2016bmv,*Cardona:2016bpi,*Cardona:2016wcr,*Gomez:2016cqb}.

The formulae for amplitudes directly obtained from ambitwistor strings are based on the scattering equations. At loop level, the formulae give the loop integrand, expressed in a manner analogous to the tree-level CHY formulae, but now based on the loop-level scattering equations \cite{Geyer:2015bja,Geyer:2015jch,Geyer:2016wjx}. A remarkable feature of these new loop-level formulae is that the representation of the loop integrand is not of the Feynman type, with ordinary propagators of the form $1/(\ell+K)^2$. Instead, the framework of the nodal Riemann sphere gives a propagator structure based mostly on linear propagators, such as $1/(2\ell\cdot K+K^2)$. This new representation discovered in \cite{Geyer:2015bja} was explored in several works, including \cite{He:2015yua,*Cachazo:2015aol,Baadsgaard:2015twa,Geyer:2015jch,Huang:2015cwh,Feng:2016nrf,Gomez:2016bmv,*Cardona:2016bpi,Geyer:2016wjx,Roehrig:2017gbt}. One conclusion is that this representation is well suited to express loop integrands in terms of higher-point tree-level amplitudes. We will see in this paper that the contributions to the $n$-particle one-loop integrand are a forward limit of contributions to an $(n+2)$-particle tree-level amplitude; the forward limit is the gluing of the 2 extra particles into a loop. This is analogous to the Feynman tree theorem. Ref.~\cite{Baadsgaard:2015twa} provides the most advanced discussion available at the moment on formal aspects of the new type of loop integrand representation, and also presents a higher-loop construction (Q-cuts) in the spirit of the Feynman tree theorem.

Let us now discuss in more detail the relationship between the scattering equations and the double copy. At tree level, the CHY formulae have been shown to provide an elegant alternative representation of the BCJ structure; see e.g.~\cite{Cachazo:2013iea,Monteiro:2013rya,Naculich:2014naa,Huang:2015yka,Bjerrum-Bohr:2016axv,Mizera:2017cqs,*Mizera:2017rqa}. In fact, the predecessor to the new developments of scattering equations and ambitwistor strings, namely Witten's twistor string theory \cite{Witten:2003nn} and the resulting `connected' formula for amplitudes \cite{Roiban:2004yf}, had already been shown to be closely intertwined with the BCJ and KLT relations \cite{Cachazo:2012uq,Cachazo:2012da}. In recent work, starting from \cite{Cachazo:2013iea}, much of the attention has focused on the remarkable properties of the CHY Pfaffian, the crucial building block that is `squared' from gauge theory to gravity. Along this line of work, Ref.~\cite{Fu:2017uzt} by Fu, Du, Huang and Feng is of particular interest to us. It presents an explicit and straightforward algorithm to write down tree-level BCJ numerators, based on a decomposition of the CHY Pfaffian. Aside from shedding light on the BCJ struture, it also provides a simple tool for dealing with the CHY amplitude expressions without explicitly solving the scattering equations. 

It is natural then to study the extension of the double-copy structure to one loop using the formalism of the scattering equations. Recently, He and Schlotterer \cite{He:2016mzd,He:2017spx} have addressed this question successfully in the case of supersymmetric theories (with various degrees of supersymmetry). They have found a natural BCJ-type structure at one loop, which relates to the tree-level structure in the same type of forward limit that we mentioned above. The details differ from the original loop-level BCJ conjecture \cite{Bern:2010ue}. The origin of this distinction is the representation of the loop momentum: the results obtained in the representation that naturally comes out of the scattering equations are not easily translated into the standard BCJ numerators associated to Feynman-type propagators. Ref.~\cite{He:2017spx} stopped short of detailing this structure in the non-supersymmetric case, because the forward-limit procedure used for the loop integrand is divergent without the cancellations provided by supersymmetry.

In this work, we find that the type of BCJ structure described in \cite{He:2016mzd,He:2017spx} also exists in the absence of supersymmetry, if one relaxes certain integrand-level identities to be valid only up to terms that vanish upon loop integration. While our results also follow from a forward limit, the procedure we apply works already at the level of the individual numerators, and therefore avoids the usual diverging propagators. In particular, we obtain one-loop BCJ numerators directly from the tree-level BCJ numerators of \cite{Fu:2017uzt} via
$$
N_{\text{1-loop}}(i_1,i_2,...,i_n) = \sum_r N_{\text{tree-level}}(+,i_1,i_2,...,i_n,-)\,,
$$
where $\pm$ represent the back-to-back loop momentum $\pm \tilde\ell$ (an on-shell version such that $\tilde \ell^2=0$), and the sum is over the gluon states running in the loop. This relation between tree-level numerators and one-loop numerators is derived from the equivalent relation between the tree-level CHY Pfaffian introduced in \cite{Cachazo:2013hca} and its one-loop analogue determined in \cite{Geyer:2015jch} from the ambitwistor string. Our results, as those of  \cite{He:2016mzd,He:2017spx}, do not directly translate into conclusions for what the BCJ structure should be in a representation with Feynman-type propagators.

This article exemplifies the remarkable properties of the novel type of representation for the loop integrand. Firstly, we easily extend a technique that was successful at tree level to loop level. Secondly, the colour-kinematics duality and the double copy are manifest. Thirdly, we are able to straightforwardly perform explicit tests of our formulae, namely to check gauge invariance and to match previously known forms of the integrand,  both of which require non-trivial procedures in a Feynman-type representation. The simplicity of these tests follows from the fact that, in the representation we use, certain integrand-level expressions are trivially recognised to vanish upon loop integration  \cite{Baadsgaard:2015twa}. We write the one-loop contribution to the scattering amplitude in the form
\begin{equation}
 {\mathcal A}^{(1)}=\int \frac{\d^D\ell}{\ell^2} \, \mathfrak{I(\ell)}\,,\label{eq:loopintgeneral}
\end{equation}
where $\mathfrak{I}$ depends on the loop momentum but not on $\ell^2$; we shall discuss this representation in the review section. Now suppose that 
\begin{equation}
\mathfrak{I'(\ell)} = \mathfrak{I(\ell)} + \Delta \mathfrak{I(\ell)}\,, \qquad \text{with}  \quad 
\Delta \mathfrak{I(\lambda\ell)} = \lambda^m \,\Delta \mathfrak{I(\ell)}\,, 
\end{equation}
for some integer $m$. It follows that, in dimensional regularisation,\footnote{This statement follows from a change of variables $\ell \to \lambda\ell$, which cannot alter the result of the integral, but leads to $\int \d^D\ell \; \Delta\mathfrak{I(\ell)}/\ell^2 \to \lambda^{D-2+m} \int \d^D\ell \; \Delta\mathfrak{I(\ell)}/\ell^2$. In dimensional regularisation, $D\neq2-m$, so the integral must vanish.}
\begin{equation}
\label{eq:Delta0}
\int \frac{\d^D\ell}{\ell^2} \, \Delta \mathfrak{I(\ell)} =0 \qquad \Longrightarrow \qquad \mathfrak{I'(\ell)} \simeq \mathfrak{I(\ell)} \,,
\end{equation}
where we use the symbol $\simeq$ to denote that both $\mathfrak{I'(\ell)}$ and $\mathfrak{I(\ell)}$ give the same result after loop integration, and are therefore valid forms of the loop integrand for the amplitude $  {\mathcal A}^{(1)}$. More generally, this is also true if $\Delta \mathfrak{I(\ell)}$ is a (finite) linear combination of terms with distinct homogeneity degree $m$. Notice that the argument still holds in the presence of the $i\epsilon$ terms that define the loop integration contour. Since a finite rescaling in $\ell$ could be accompanied by a rescaling of the $\epsilon$'s, the argument above still applies. We assume explicitly that an $i\epsilon$ prescription exists for the new type of representation of the loop integrand. A natural prescription was proposed and checked in simple examples in \cite{Baadsgaard:2015twa}, and it would be interesting to investigate its origin in the ambitwistor string, but we will not be concerned with this here.

The paper is organised as follows. We start in section~\ref{sec:review} with a review of the BCJ and CHY structures at tree level and at one loop, including the one-loop CHY-type formulas derived from ambitwistor strings, which are the main ingredient for our results. In section~\ref{sec:nums}, we present the algorithm to obtain one-loop BCJ numerators and give various examples. We describe several tests of the results for Yang-Mills theory in section~\ref{sec:YM}, while also giving an explicit example of how to construct the loop integrand from the BCJ numerators. In section~\ref{sec:doublecopy}, we discuss the double copy to gravity, leading to formulae for both NS-NS gravity and pure gravity. To conclude, we summarise our results and comment on open questions in section~\ref{sec:discussion}.


\section{Review} \label{sec:review}

We start this section with brief reviews of the CHY representation of tree-level scattering amplitudes and of the BCJ colour-kinematics duality and double copy. The models underlying the simple and compact CHY formulae for amplitudes --  two-dimensional chiral conformal field theories, known as ambitwistor strings \cite{Mason:2013sva} -- naturally extend the scattering equation formalism to higher loop order. We review key formulae, as well as the specific representation of the loop integrand these models result in.

\subsection{Tree-level amplitudes and BCJ numerators}
\paragraph{The CHY representation of tree-level amplitudes.} 

The Cachazo-He-Yuan (CHY) representation \cite{Cachazo:2013gna, Cachazo:2013hca, Cachazo:2013iea} expresses tree-level scattering amplitudes for $n$ massless particles in $D$ dimensions as integrals over the moduli space of a punctured Riemann sphere $\mathfrak{M}_{0,n}$,
\begin{equation}
 {\mathcal{A}}^{(0)}_n=\int_{\mathfrak{M}_{0,n}}\!\!\d\mu_{0,n}\,\cI\,,\qquad\qquad \d\mu_{0,n}\equiv \frac{\prod_{i=1}^n \d\sigma_i}{\vol\SL(2,\C)} \ \prod_{i=1}^n{}' \, \delta\Bigg(\sum_{j\neq i} \frac{k_i\cdot k_j}{\sigma_{ij}}\Bigg)\,,
\end{equation}
with $\sigma_i\in\CP^1$ and $\sigma_{ij}=\sigma_i-\sigma_j$. The measure $\d\mu_{0,n}$ is universal to all massless scattering amplitudes and fully localises the moduli space integral onto the solutions of the so-called {\it scattering equations}\footnote{These scattering equations also feature prominently in the high-energy scattering of strings \cite{Gross:1987kza}. See also \cite{Casali:2014hfa} for more details on the relation to ambitwistor strings and the CHY formulae.} \cite{Cachazo:2013gna, Cachazo:2013hca}
\begin{equation}
 E_i\equiv\sum_{j\neq i} \frac{k_i\cdot k_j}{\sigma_{ij}}=0\,.
\label{eq:SE}
\end{equation}
Both the scattering equations and the (theory-specific) integrand $\cI$ transform covariantly under M\"obius transformations, in such a manner that the amplitude is invariant. Fixing this redundancy introduces the usual Jacobian and removes three redundant scattering equations (hence the prime in the formula). This leaves exactly $\mathrm{dim}_{\mathbb C}(\mathfrak{M}_{0,n})=n-3$ constraints, which fully localise the integral.

In all cases of interest studied so far, the CHY integrand $\cI$ exhibits a double-copy structure, 
\begin{equation}
\cI=\cI_L\,\cI_R\,,
\end{equation}
where the factors $\cI_{L,R}$ depend explicitly on the marked points $\sigma_i$ and on the scattering data, e.g.~the null momenta $k_i$, the polarisation vectors $\epsilon_i$, or the SU$(N_c)$ colour indices $a_i$.\footnote{Notice that the factorisation poles of the amplitudes are due to poles in $\sigma_{ij}$ of the integrand $\cI$, evaluated on solutions to the scattering equations; see e.g. \cite{Dolan:2013isa}. The explicit dependence of $\cI$ on the Mandelstam variables is polynomial.} A theory is thus specified by its building blocks $\cI_{L}$ and $\cI_R$. While CHY representations have been given for a wide family of theories \cite{Cachazo:2014nsa, Cachazo:2014xea}, ranging from Einstein-Yang-Mills to effective scalar theories such as Born-Infeld or the non-linear sigma model, we focus here on the formulae for gravity and Yang-Mills theory:
\begin{align}
 \cI_{\text{YM}}=\cI_{\text{SU}(N_c)}\,\cI_{\text{kin}}\,,\qquad\qquad \cI_{\text{grav}}=\cI_{\text{kin}}\,\widetilde\cI_{\text{kin}}\,.\label{eq:CHY_YM_grav}
\end{align}
where the `colour' and `kinematic' integrand factors are given by
\begin{subequations}
\begin{align}
 &\cI_{\text{SU}(N_c)}=\sum_{\rho\in {S}_{n}/\Z_n}\frac{\tr\left(T^{\rho(a_1)}...T^{\rho(a_n)}\right)}{\sigma_{\rho(a_1)\rho(a_2)}...\sigma_{\rho(a_n)\rho(a_1)}}\,,\label{equ:Icoltree}\\
 &\cI_{\text{kin}}=\pf'(M)\equiv\frac{(-1)^{\hat i+\hat j}}{\sigma_{\hat i\hat j}}\pf\left(M^{\hat i \hat j}_{\hat i \hat j}\right)\,, \qquad \widetilde\cI_{\text{kin}}=\cI_{\text{kin}}(\epsilon_i\to\tilde\epsilon_i)\,.
\label{equ:Ikintree}
\end{align}
\end{subequations}
The sum in $\cI_{\text{SU}(N_c)}$ runs over non-cyclic permutations, denoted by ${S}_{n}/\Z_n$. The $2n\times 2n$ antisymmetric matrix $M(\{k_i,\epsilon_i,\sigma_i\})$ 
defining $\cI_{\text{kin}}$ is determined by  
 \begin{subequations}\label{eq2:def_M_CHY}
 \begin{align} 
  & &&M=\begin{pmatrix} A & -C^T \\ C & B \end{pmatrix}\,, &&\\
  &A_{ij}=\frac{k_i\cdot k_j}{\sigma_{ij}}\,,&&B_{ij}=\frac{\epsilon_i\cdot \epsilon_j}{\sigma_{ij}}\,, &&C_{ij}=\frac{\epsilon_i\cdot k_j}{\sigma_{ij}}\,,\\
  &A_{ii}=0\,, && B_{ii}=0\,, && C_{ii}=-\sum_{j\neq i}C_{ij}\,.
 \end{align}
 \end{subequations}
On solutions to the scattering equations \eqref{eq:SE}, $M$ has co-rank two,\footnote{Its kernel is spanned by the vectors $(1,\dots,1,0,\dots,0)$ and $(\sigma_1,\dots,\sigma_n,0,\dots,0)$.} and thus $\pf(M)=0$. However, an invariant quantity, the {\it reduced} Pfaffian $\pf'(M)$, can be defined by removing any two rows and columns $\hat i$ and $\hat j$ such that $1\leq \hat i < \hat j \leq n$, leading to $\cI_{\text{kin}}$ as given in \cref{equ:Ikintree}.

The dependence of $\cI_{\text{grav}}$ on two sets of polarisation vectors indicates that the scattering states have the polarisation tensors $\varepsilon_i^{\mu\nu}=\epsilon_i^\mu\tilde\epsilon_i^\nu$. The gravity theory described in this way is the theory of NS-NS gravity (the name is imported from string theory), describing gravitons, dilatons and (2-form) B-field states. At tree level, if the external states are restricted to describing gravitons (an appropriate symmetric and traceless linear combination of terms $\epsilon_i^\mu\tilde\epsilon_i^\nu$), then the scattering amplitudes coincide with those of pure Einstein gravity.

\paragraph{The colour-kinematics duality.} 
The splitting of the CHY integrands into colour and kinematic factors in \cref{eq:CHY_YM_grav} is highly suggestive of the Bern-Carrasco-Johansson (BCJ) colour-kinematics duality for gauge theory and the associated double copy to gravity \cite{Bern:2008qj}. To make the connection between the two more explicit, 
 we should relate the CHY framework to the BCJ Feynman-like diagrams.

Our starting point is an expansion of the Yang-Mills amplitude into cubic diagrams $\Gamma_\alpha$, 
\begin{equation}
\label{equ:cubicYMtree}
 \mathcal{A}^{(0)}_{\text{YM}}=\sum_{\Gamma_\alpha} \frac{c_\alpha\, N_\alpha}{D_\alpha}\,,
\end{equation}
where each term contributes with a colour factor $c_\alpha$, a kinematic numerator $N_\alpha$ dependent on the external momenta and on the gluon polarisations,\footnote{In this paper, we will only consider the scattering of gluons. More generally, we could also consider matter states, e.g.~in supersymmetric gauge theories.} and a product $1/D_\alpha$ of scalar propagators of the cubic graph. The colour factors are composed of the structure constants $f^{abc}$ of the Yang-Mills Lie algebra, and therefore obey Jacobi relations
\begin{equation}
 c_\alpha \pm c_\beta \pm c_\gamma=0\,, \label{equ:BCJtripletcol}
\end{equation}
for suitable choices of $(\alpha, \beta, \gamma)$ and signs. The colour-kinematics duality \cite{Bern:2008qj} then ascertains that there exists a dual set of kinematic numerators $N_\alpha$ satisfying the same relations,
\begin{equation}
 N_\alpha \pm N_\beta \pm N_\gamma=0\,.\label{equ:BCJtripletnum}
\end{equation}
Kinematic numerators satisfying \cref{equ:BCJtripletnum} whenever the corresponding colour factors satisfy  \cref{equ:BCJtripletcol} are known as BCJ numerators, and will be the central object of interest throughout this article. An important point is that this set of numerators is non-unique, i.e.~there exist different choices giving the same amplitude. We will see later that this statement also applies at one loop. The loop-level conjecture of the colour-kinematics duality is the straightforward extension of the statement above, but applied to the loop integrand expressed in terms of cubic diagrams, where the kinematic numerators and the propagators depend on the loop momenta \cite{Bern:2010ue}.

Given a set of BCJ numerators from Yang-Mills theory, a gravity amplitude is obtained straightforwardly via the BCJ double copy \cite{Bern:2008qj}: we substitute the colour factors in \eqref{equ:cubicYMtree} by another set of Yang-Mills numerators,
\begin{equation}
 \mathcal{A}^{(0)}_{\text{grav}}=\sum_{\Gamma_\alpha} \frac{N_\alpha\, \tilde N_\alpha}{D_\alpha} 
\,, \qquad \tilde N_\alpha= N_\alpha(\epsilon_i\to\tilde\epsilon_i) \,.
\end{equation}
The external states in the gravity amplitude have the polarisation tensors $\varepsilon_i^{\mu\nu}=\epsilon_i^\mu\tilde\epsilon_i^\nu$, exactly as we discussed before for the corresponding CHY formula.

We can consider the BCJ structure starting from an alternative decomposition of the colour dependence of the gauge theory amplitude, in terms of colour traces,
\begin{equation}
 \mathcal{A}^{(0)}_{\text{YM}}=\sum_{\rho\in {S}_{n}/\Z_n} A_{\text{YM}}(\rho(a_1),...,\rho(a_n)) \ \tr\left(T^{\rho(a_1)}...T^{\rho(a_n)}\right)\,.
\end{equation}
The existence of numerator relations \eqref{equ:BCJtripletnum} is then equivalent to the fact that the partial amplitudes $A_{\text{YM}}$ satisfy the BCJ relations \cite{Bern:2008qj},\footnote{These relations appear in string theory as monodromy relations \cite{BjerrumBohr:2009rd,*Stieberger:2009hq}.}
\begin{equation}
 \sum_{j=2}^n k_1\cdot k_{23...j} \,A_{\text{YM}}\big(2, 3, . . ., j, 1, j+1, . . ., n\big) = 0\,, \qquad  k_{i\ldots j} = \sum_{a=i}^j k_a\,.
\end{equation}
In the CHY framework, these relations emerge elegantly as relations among the Parke-Taylor factors appearing in \eqref{equ:Icoltree} \cite{Cachazo:2012uq,Cachazo:2013gna},
\begin{equation}
 \sum_{j=2}^n \frac{k_1\cdot k_{23...j} }{\sigma_{23}...\sigma_{j1}\sigma_{1\,j+1}...\sigma_{n2}} = 0\,,\qquad \text{mod }E_i\,,
\end{equation}
that is, on the support of the scattering equations \eqref{eq:SE}. This short and elegant implementation of the BCJ relations should serve as a first indication that the BCJ numerators $N_\alpha$ can also be derived effectively from the scattering-equations formalism. \\

\begin{figure}[ht]
\begin{center}
 \begin{tikzpicture}[scale=1]
 \draw (0,0) -- (2.25,0) ;
 \draw[dotted, thick] (2.25,0) -- (3.25,0) ;
 \draw (3.25,0) -- (4.5,0);
 \draw (1,0) -- (1,1) ;
 \draw (2,0) -- (2,1) ;
 \draw (3.5,0) -- (3.5,1) ;
 \node at (-0.4,0) {$1$};
 \node at (4.9,0) {$n$};
 \node at (1,1.4) {$\rho(2)$};
 \node at (2,1.4) {$\rho(3)$};
 \node at (3.5,1.4) {$\rho(n-1)$};
\end{tikzpicture}
\end{center}
\caption{The tree-level BCJ master numerators $N\big(1,\rho(2,...,n-1),n\big)$, correspond to half-ladder diagrams with legs $1$ and $n$ at opposite endpoints. }
\label{fig:half-ladder}
\end{figure}
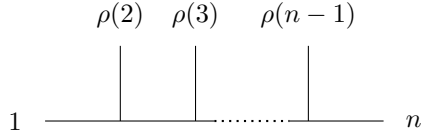

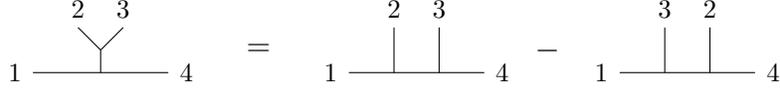
\begin{figure}[ht]
\begin{center}
 \begin{tikzpicture}[scale=0.6]
 \draw (0,0) -- (3,0) ;
 \draw (1,0) -- (1,1) ;
 \draw (2,0) -- (2,1) ;
 \node at (-0.4,0) {$1$};
 \node at (3.4,0) {$4$};
 \node at (1,1.4) {$2$};
 \node at (2,1.4) {$3$};
 \node at (4.4,0.5) {{\large $-$}};
 \draw (6,0) -- (9,0) ;
 \draw (7,0) -- (7,1) ;
 \draw (8,0) -- (8,1) ;
 \node at (5.6,0) {$1$};
 \node at (9.4,0) {$4$};
 \node at (7,1.4) {$3$};
 \node at (8,1.4) {$2$};
 \node at (-2,0.5) {{\large $=$}};
 \draw (-7,0) -- (-4,0) ;
 \draw (-5.5,0) -- (-5.5,0.5) ;
 \draw (-5.5,0.5) -- (-5,1) ;
 \draw (-5.5,0.5) -- (-6,1) ;
 \node at (-7.4,0) {$1$};
 \node at (-3.6,0) {$4$};
 \node at (-6,1.4) {$2$};
 \node at (-5,1.4) {$3$};
\end{tikzpicture}
\end{center}
\caption{Example at 4 points of how a diagram relates to master diagrams via Jacobi relations. This equation applies to both the colour factors and (by definition) the BCJ numerators associated to the diagrams.}
\label{fig:jacobiexample}
\end{figure}

The principal idea for deriving numerators satisfying \cref{equ:BCJtripletnum} from the CHY formulae, following \cite{Fu:2017uzt} and building on earlier work \cite{Cachazo:2013iea,BjerrumBohr:2010hn,Fu:2012uy}, is to expand both the Yang-Mills amplitude and the gravity amplitude in a Dixon-Del Duca-Maltoni (DDM) half-ladder basis \cite{DelDuca:1999rs}, 
\begin{subequations}
\begin{align}
 \mathcal{A}_{\text{YM}}&=\sum_{\rho\in S_{n-2}} c(1,\rho,n) \,A_{\text{YM}}(1,\rho,n)\,, \\
 \mathcal{A}_{\text{grav}}&=\sum_{\rho\in S_{n-2}} N(1,\rho,n) \,A_{\text{YM}}(1,\rho,n)\,,\label{equ:decMintoA}
\end{align}
\end{subequations}
where both the colour factors
$$
c(1,\rho,n) = f^{a_1a_{\rho(2)}b_1} \, f^{b_1a_{\rho(3)}b_2} \cdots f^{b_{n-3}a_{\rho(n-1)}a_n}\,, \qquad  \text{with} \quad [T^a,T^b]=f^{abc} T^c \,,
$$
and the kinematic numerators $N(1,\rho,n)$ are associated with cubic diagrams forming a `half-ladder'; see \cref{fig:half-ladder}.
If this can be achieved, the amplitudes satisfy the double-copy structure, since the gravity amplitude $\mathcal{A}_{\text{grav}}$ is constructed from the gauge theory amplitude $\mathcal{A}_{\text{YM}}$ by replacing the colour factors $c(1,\rho,n)$ by kinematic numerators $N(1,\rho,n)$. The numerators of DDM half-ladder diagrams are known as {\it master numerators}, since the numerators of any other cubic diagram can be obtained from these via the Jacobi-type relations \eqref{equ:BCJtripletnum}; this mirrors the analogous statement for colour factors, as in \cref{fig:jacobiexample}. The colour-kinematics duality is thus satisfied by construction.

The challenge is then to decompose the gravity amplitude into Yang-Mills amplitudes \eqref{equ:decMintoA}. In the CHY language, this can be achieved by rewriting the `colour' \eqref{equ:Icoltree} and `kinematic' \eqref{equ:Ikintree} integrand factors \cite{Cachazo:2013iea},\footnote{The so-called KLT orthogonality guarantees that such an expansion exists, \cite{Cachazo:2013gna, Cachazo:2012da}. A prior version of this type of decomposition, valid for a (pure spinor) supersymmetrised version of the Pfaffian, was given in \cite{Mafra:2011kj}.}
\begin{subequations}
\begin{align}
& \cI_{\text{SU}(N_c)} \; = \sum_{\rho\in S_{n-2}} 
\frac{ c(1,\rho,n) 
}{\sigma_{1\rho(2)}\,\sigma_{\rho(2)\rho(3)} \cdots \sigma_{\rho(n-1)n} \,\sigma_{n1}} \,,  \\
& \cI_{\text{kin}} \; =\; \mathrm{Pf} \, ' (M)
\; = \sum_{\rho\in S_{n-2}} 
\frac{ N(1,\rho,n)
}{\sigma_{1\rho(2)}\,\sigma_{\rho(2)\rho(3)} \cdots \sigma_{\rho(n-1)n} \,\sigma_{n1}} \,,
\qquad \text{mod }E_i\,.
\end{align}
\end{subequations}
Notice that neither the colour factors $c(1,\rho,n)$ nor the kinematic numerators $N(1,\rho,n)$ depend on the punctures $\sigma_i$. While there exist several distinct choices of numerators achieving this goal, we will focus our attention on a recent construction of \cite{Fu:2017uzt}; see also the closely related work \cite{Chiodaroli:2017ngp}. By definition, the Pfaffian is a product over closed cycles, and this structure can be exploited to find a convenient Ansatz for a recursive expansion. In \cite{Fu:2017uzt}, Fu, Du, Huang and Feng demonstrate beautifully that the parameters in such an Ansatz are completely fixed by the requirement of gauge invariance, and derive a surprisingly simple algorithm for all tree-level BCJ numerators. Instead of reviewing here their tree-level algorithm, we will describe it in our one-loop construction in \cref{sec:nums}.

\subsection{One-loop integrands from ambitwistor strings}
The existence of the remarkably simple CHY representation of the S-matrix is elegantly explained by ambitwistor strings \cite{Mason:2013sva, Ohmori:2015sha, Berkovits:2013xba} -- chiral worldsheet models with an auxiliary target space, the space of complex null geodesics. Ambitwistor string correlation functions give rise to the CHY representation of scattering amplitudes, and the physical understanding provided by the underlying CFT has been crucial in extending the scattering equations formalism to loop level \cite{Adamo:2013tsa,Geyer:2015bja,Geyer:2015jch,Geyer:2016wjx}. 

In the genus expansion of the ambitwistor string, $g$-loop amplitudes are represented by integrals over the moduli space of punctured genus-$g$ Riemann surfaces $\mathfrak{M}_{g,n}$. Specialising to one loop, the loop integrand\footnote{The full amplitude moreover involves an integration over the loop momentum $\ell$.} is thus given by an integral over the punctures, as well as the fundamental domain of the modular parameter $\tau$ of the torus. While conceptually simple, the mathematical framework on higher genus Riemann surfaces is computationally challenging and obscures the relative simplicity of the expected loop integrand. 

Due to modular invariance, however, the support of the one-loop scattering equations localises the $\tau$ integral  on the cusp or non-separating degeneration $\tau\rightarrow i\infty$, \cite{Geyer:2015bja, Geyer:2015jch}. On the resulting nodal Riemann sphere, one-loop amplitudes take the following form:
\begin{equation}
{\mathcal A}^{(1)}_n=\int \frac{\d^D\ell}{\ell^2}\int_{\mathfrak{M}_{0,n+2}}\hspace{-20pt}\ \d\mu_{1,n}^{(\text{nod})}\,\cI^{(1)}\,,\qquad\qquad \d\mu_{1,n}^{(\text{nod})}\equiv\frac{\prod_{a} \d\sigma_a}{\vol\SL(2,\C)}\prod_{a}{}' \delta\Big(E_a^{(\text{nod})}\Big)\,.\label{eq:loopint}
\end{equation}
In \cref{eq:loopint}, $a$ runs over all external particles as well as the two parameters describing the node, $\sigma_+$ and $\sigma_-$, corresponding to the insertions of $+\ell$ and $-\ell$ respectively. The nodal scattering equations $E_a^{(\text{nod})}$ in the measure $\d\mu_{1,n}^{(\text{nod})}$ bear a remarkable similarity to the tree-level scattering equations for $n+2$ particles,
\begin{subequations}\label{equ:nodalSE}
\begin{align}
 E_+^{(\text{nod})}&\equiv \sum_{i} \frac{\ell\cdot k_i}{\sigma_{+i}}\,,\\
 E_-^{(\text{nod})}&\equiv -\sum_{i} \frac{\ell\cdot k_i}{\sigma_{-i}}\,,\\
 E_i^{(\text{nod})}&\equiv\frac{k_i\cdot \ell}{\sigma_{i+}}-\frac{k_i\cdot \ell}{\sigma_{i-}} +\sum_{j\neq i} \frac{k_i\cdot k_j}{\sigma_{ij}}\,.
\end{align}
\end{subequations}
In fact, the nodal measure may be written compactly as 
\begin{equation}
 \d\mu_{1,n}^{(\text{nod})}= \d\mu_{0,n+2}\big|_{\tilde{\ell}^2=0}\,,\label{equ:1-loop-measure}
\end{equation}
with two additional particles with back-to-back on-shell momenta $\pm\tilde\ell$. This on-shell momentum $\tilde\ell$ relates to the loop momentum $\ell$ as $\tilde\ell=\ell+\eta$, where $\eta$ satisfies $\ell\cdot\eta=k_i\cdot\eta=\epsilon_i\cdot\eta=0$ and $\eta^2=-\ell^2$; so $\eta$ can be thought of as a higher-dimensional contribution. Therefore, if $\cI^{(1)}$ depends on the loop momentum only via $\ell\cdot k_i$ and $\ell\cdot \epsilon_i$, we have $\cI^{(1)}(\ell)=\cI^{(1)}(\tilde\ell)$. We will see below that this interpretation in terms of $n+2$ on-shell particles has immediate consequences for the representation of the loop integrand, and plays an important role in the construction of the BCJ numerators.

It is worth highlighting that while the genus-one representation of the amplitude has so far only been achieved for type II supergravity (sugra) and only in the critical dimension $D=10$, the one-loop formalism based on the nodal Riemann sphere can be applied to a variety of theories in any dimension $D$. In fact, this formalism was used in \cite{Geyer:2015bja} for super-Yang-Mills (sYM) theory and in \cite{Geyer:2015jch} for gravity and gauge theory in a variety of dimensions and degrees of supersymmetry, even though a genus-one ambitwistor string representation is not known (or perhaps even expected to exist) for such theories. Moreover, there is strong evidence for the validity of its higher-genus extension \cite{Geyer:2016wjx}, representing $g$-loop amplitudes as integrals over the moduli space of punctured $g$-nodal Riemann spheres. 

Similarly to tree level, there is a factorisation\, $\cI^{(1)}=\cI^{(1)}_L\,\cI^{(1)}_R$\, in the theories of interest that have been explored so far. The fundamental building blocks for the one-loop integrands of Yang-Mills theory and gravity are given by the following colour and kinematic integrand factors \cite{Geyer:2015jch}
\begin{subequations}
\begin{align}
 \cI^{(1)}_{\text{SU}(N_c)} &=\sum_{\rho\in S_n}\frac{
\tr\left(T^{\rho(a_1)}...T^{\rho(a_n)}\right)}{\sigma_{+\,\rho(1)}\sigma_{\rho(1)\,\rho(2)}\dots\sigma_{\rho(n)\,-}\sigma_{-\,+}}\, + \, \text{double-trace terms}\,,\\
 \cI^{(1)}_{\text{susy}} &= \cI^{(1)}_{\text{NS}}+ \cI^{(1)}_{\text{R}}\,.
\end{align}
\end{subequations}
The ambitwistor-string origin of these expressions allows for the identification of contributions from individual GSO sectors, as well as the scalar contribution (following a similar analysis in the string literature \cite{Tourkine:2012vx}),
 \begin{align}
 \cI^{(1)}_{\text{NS}} =\sum_r\pf'\big(M_{\text{NS}}^r\big)\,, \qquad \cI^{(1)}_{\text{R}} =-\frac{c_D}{(\sigma_{+-})^2}\ \pf\big(M_2\big)\,,\qquad
\cI^{(1)}_{\text{scal}} =\frac{1}{(\sigma_{+-})^2}\ \pf\big(M_3\big)\,.
 \end{align}
Here, $c_D$ is a dimension-dependent constant\footnote{The dimensional reduction of 8 Majorana-Weyl spinors in $D=10$ leads to $c_{10} = 8$, $c_8 = 8$, $c_6 = 2$ and $c_4 = 2$.}, and the matrices $M_{\alpha}$ for $\alpha=2,3$ (stemming from the different spin structures $(0,0)$ and $(1,0)$) are defined similarly to \cref{eq2:def_M_CHY} by
 \begin{subequations}
 \begin{align} \label{eq2:def_Malpha}
  & && M_{\alpha}=\begin{pmatrix} A & -C^T \\ C & B \end{pmatrix}\,, &&\\
  &A_{ij}=k_i\cdot k_j\, S_{\alpha}(\sigma_{ij})\,,&&B_{ij}=\epsilon_i\cdot \epsilon_j \,S_{\alpha}(\sigma_{ij})\,, &&C_{ij}=\epsilon_i\cdot k_j \,S_{\alpha}(\sigma_{ij})\,,\\
  &A_{ii}=0\,, && B_{ii}=0\,, && C_{ii}=-\sum_{j\neq i}\frac{\epsilon_i\cdot k_j}{\sigma_{ij}}-\frac{\epsilon_i\cdot \ell}{\sigma_{i+}}+\frac{\epsilon_i\cdot \ell}{\sigma_{i-}}\,.
 \end{align}
 \end{subequations}
with $S_3=\sigma_{ij}^{-1}$ and $S_2=\sigma_{ij}^{-1}\left(\sqrt{\frac{\sigma_{i+}\sigma_{j-}}{\sigma_{i-}\sigma_{j+}}}+\sqrt{\frac{\sigma_{i-}\sigma_{j+}}{\sigma_{i+}\sigma_{j-}}}\right)$. As discussed above, these integrands only depend on the loop momentum via $\ell\cdot k_i$ and $\ell\cdot \epsilon_i$, and hence $\mathcal{I}^{(1)}(\ell)=\mathcal{I}^{(1)}(\tilde\ell)$. The NS integrand $\mathcal{I}_{\text{NS}}^{(1)}$ therefore  manifests the interpretation of adding two additional particles with back-to-back momenta $\pm\tilde\ell$ with $\tilde\ell=\ell+\eta$,
\begin{equation}\label{eq:MNS}
 M_{\text{NS}}^r = M_{n+2}^{\text{tree}}\Bigg|_{\,\tilde{\ell}^2=0\,,\;\epsilon_{+}=\epsilon^r\,,\;\epsilon_{-}=(\epsilon^r)^\dagger}\,.
\end{equation}
and the sum runs over a basis $\epsilon^r$ of polarisation vectors for the two additional particles. The completeness relation for this basis, defined via $\epsilon^r\cdot q=0$ for a null $q$, is
\begin{equation}
\sum_{r=1}^{D-2} \epsilon^r_\mu\, (\epsilon^r)^\dagger_\nu \,=\, 
\eta_{\mu\nu} - \frac{k_\mu q_\nu+k_\nu q_\mu}{k\cdot q}
\,\equiv\, \Delta_{\mu\nu} \,.
\end{equation}
We can produce the substitutions
\begin{equation}
\label{equ:looppolsub}
\sum_{r} \epsilon_+\cdot\epsilon_- \,=\, D-2 \,, \qquad \sum_{r} (\epsilon_+\cdot v)\,(\epsilon_-\cdot w) \,=\, \Delta_{\mu\nu} v^\mu w^\nu   \ \leadsto\ v\cdot w \,,
\end{equation}
where in the second equation we are allowed to drop the $q$-dependent term in $\Delta_{\mu\nu}$, since its contribution vanishes on the solutions to the scattering equations; see \cite{Roehrig:2017gbt}. These substitutions will later allow us to obtain $n$-particle one-loop BCJ numerators from ($n+2$)-particle tree-level BCJ numerators.

As at tree level, the Yang-Mills and gravity integrands, given here with and without supersymmetry, exhibit the double-copy relation:
\begin{subequations}  \label{equ:loop-integrands}
\begin{align}
 \cI^{(1)}_{\text{sYM}} &=\cI^{(1)}_{\text{SU}(N_c)}\,\cI^{(1)}_{\text{susy}}\,, & \cI^{(1)}_{\text{sugra}} &=\cI^{(1)}_{\text{susy}}\,\widetilde\cI^{(1)}_{\text{susy}}\,,\\
 \cI^{(1)}_{\text{YM}} &=\cI^{(1)}_{\text{SU}(N_c)}\,\cI^{(1)}_{\text{NS}}\,,  & \cI^{(1)}_{\text{NS-NS}} &=\cI^{(1)}_{\text{NS}}\,\widetilde\cI^{(1)}_{\text{NS}}\,.
 \end{align}
\end{subequations}
In the case of NS-NS gravity, we have the graviton, dilaton and B-field states running in the loop. If we want to obtain pure gravity, with only the graviton states in the loop, we must subtract the dilaton and B-field contributions. This is particularly easy in $D=4$, since the B-field has a single degree of freedom, the axion, and therefore it suffices to subtract the contribution of two scalars running in the loop, 
\begin{equation}
\label{equ:loop-integrand-pure}
\cI^{(1)}_{\text{4D-pure-grav}} =\cI^{(1)}_{\text{NS}}\,\widetilde\cI^{(1)}_{\text{NS}} - 2\, \cI^{(1)}_{\text{scal}}\, \widetilde\cI^{(1)}_{\text{scal}} \,.
\end{equation}

\paragraph{New representation of the loop integrand.} One feature of this scattering-equations based representation particularly worth highlighting is that it gives rise to a non-standard representation of the loop integrand:\footnote{In fact, this can already be observed from the general form of the amplitude given in \cref{eq:loopint} and \cref{equ:nodalSE} by simply counting powers of the loop-momentum. Since all scattering equations in \cref{equ:nodalSE} are linear in $\ell$, all propagators will be as well, and the only quadratic power comes from the overall $\ell^{-2}$. See \cite{Dolan:2013isa} at tree level and \cite{Geyer:2015jch} at one loop for details on this argument using factorisation.} after carrying out the integration over the moduli space $\mathfrak{M}_{0,n+2}$, the integrand contains poles linear in $\ell$ not immediately recognisable as the conventional loop propagators $(\ell+K)^2$. 

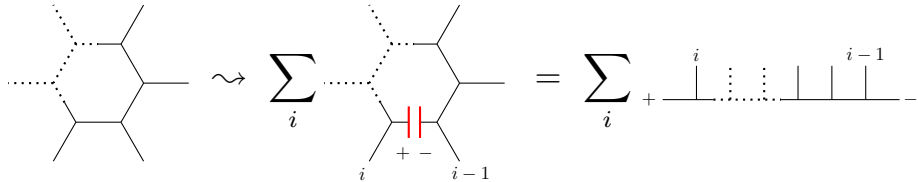
\begin{figure}[ht]\vspace{7pt}
\begin{center}
 \begin{tikzpicture}[scale=0.3]
 \draw (-0.5,0.866) -- (0,0) -- (1,0) -- (2,0) -- (3,1.732) -- (2,3.464) -- (1,3.464);
 \draw[dotted, thick] (-0.5,0.866) -- (-1,1.732) -- (0,3.464) -- (1,3.464);
 \draw (0,0) -- (-1,-1.732);
 \draw[dotted, thick] (-1,1.732) -- (-3,1.732);
 \draw[dotted, thick] (0,3.464) -- (-1, 5.196);
 \draw (3,1.732) -- (5,1.732);
 \draw (2,0) -- (3,-1.732);
 \draw (2,3.464) -- (3, 5.196);
 
 \node at (8.4,1.4) {\scalebox{1.3}{$\leadsto\,\,\displaystyle\sum_i$}};
 \draw (13.5,0.866) -- (14,0) -- (14.75,0);
 \draw (15.25,0) -- (16,0) -- (17,1.732) -- (16,3.464) -- (15,3.464);
 \draw[dotted, thick] (13.5,0.866) -- (13,1.732) -- (14,3.464) -- (15,3.464);
 \draw (14,0) -- (13,-1.732);
 \draw[dotted, thick] (13,1.732) -- (11,1.732);
 \draw[dotted, thick] (14,3.464) -- (13, 5.196);
 \draw (17,1.732) -- (19,1.732);
 \draw (16,0) -- (17,-1.732);
 \draw (16,3.464) -- (17, 5.196);
 \draw[red,thick] (14.75,-0.7) -- (14.75,0.7);
 \draw[red,thick] (15.25,-0.7) -- (15.25,0.7);
 \node at (14.5, -1.4) {\scalebox{0.7}{$+$}};
 \node at (15.5,-1.4) {\scalebox{0.7}{$-$}};
 \node at (12.7,-2.3) {\scalebox{0.7}{$i$}};
 \node at (17.5,-2.3) {\scalebox{0.7}{$i-1$}};
 
 \node at (22.5,1.4) {\scalebox{1.3}{$=\,\,\displaystyle\sum_i$}};
 \draw (26,1) -- (28.25,1);
 \draw (27.5,1) -- (27.5,2.5);
 \draw[dotted,thick] (28.25,1) -- (31.25,1);
 \draw (31.25,1) -- (36.5,1);
 \draw[dotted,thick] (29,1) -- (29,2.5);
 \draw[dotted,thick] (30.5,1) -- (30.5,2.5);
 \draw (32,1) -- (32,2.5);
 \draw (33.5,1) -- (33.5,2.5);
 \draw (35,1) -- (35,2.5);
 \node at (25.4, 1) {\scalebox{0.7}{$+$}};
 \node at (37,1) {\scalebox{0.7}{$-$}};
 \node at (27.5,3) {\scalebox{0.7}{$i$}};
 \node at (35,3) {\scalebox{0.7}{$i-1$}};
\end{tikzpicture}
\end{center}\vspace{-20pt}
\caption{Diagrammatic depiction: interpretation of the $\mathfrak{I}$ representation of loop integrands as $(n+2)$-particle tree diagrams.}
\label{fig:loopstotrees}
\end{figure}

However, there is a simple prescription to obtain such a representation from a standard loop integrand. The method is particularly easy to illustrate for numerators independent of $\ell^2$ \cite{Geyer:2015bja}. In that case, a loop integrand representation $I_{lin}$ of the type appearing in \cref{eq:loopint} can be derived from a standard `quadratic' representation $I_{qdr}$ via repeated partial fraction identities of the form
\begin{equation}
 \frac{1}{\prod_a D_a}=\sum_a \frac{1}{D_a\prod_{b\neq a}(D_b-D_a)}\,,\qquad \text{where } D_a=(\ell+K_a)^2\text{ and }K_a=\sum_{i\in I_a}k_i\,,
\label{eq:partialfrac}
\end{equation}
and shifts in the loop momentum $\ell\rightarrow \ell-K_a$ (differently for each term in the partial fraction expansion), to obtain the overall factor $\ell^{-2}$. These partial fraction identities are in fact a special case of a more widely applicable contour integral argument \cite{Baadsgaard:2015twa}: to relate the different representations for general numerators\footnote{For readability, we keep the dependence on $\ell^2$ explicit. By the first argument in $N(\ell,\ell^2)$ we mean the $\ell$ dependence of the type $\ell\cdot k_i$ or $\ell\cdot \epsilon_i$.} $N(\ell, \ell^2)$, we shift the loop momentum in the standard representation $I_{qdr}$ by $\ell\rightarrow \tilde\ell=\ell+\eta$ (c.f. \cref{eq:MNS} and \cref{equ:1-loop-measure}), where $\ell\cdot\eta=k_i\cdot\eta=\epsilon_i\cdot\eta=0$, such that the Lorentz invariants are unaffected except for $\ell^2\rightarrow \ell^2+\eta^2\equiv\ell^2+\zeta$. The integrand is then naturally written as the residue at $\zeta=0$. Applying the Cauchy residue theorem, and shifting individual terms by $\ell\rightarrow \ell-K_a$ as above, then yields $I_{lin}$:
\begin{equation}\label{eq:IlinQ}
 I_{qdr}=\sum_\Gamma\frac{N\big(\ell,\ell^2\big)}{\prod_{a\in\Gamma}D_a} \quad \leadsto \quad
 I_{lin}=\frac{1}{\ell^2}\sum_\Gamma\sum_{a\in\Gamma}\frac{N\big(\ell-K_a,\,-2\ell\cdot K_a +K_a^2\big)}{\prod_{b\neq a}(D_b-D_a)\Big|_{\ell\rightarrow \ell-K_a}} =\frac{1}{\ell^2} \ {\mathfrak I} \,, 
\end{equation}
up to terms integrating to zero in dimensional regularisation; hence ${\mathfrak I}$ does not depend on $\ell^2$. It is easy to check that the denominator factors in ${\mathfrak I}$ take the form $2\ell\cdot K+K^2$, where $K$ is a sum of external momenta. The sum over different propagators $a$ on the right-hand side of \cref{eq:IlinQ} has an intuitive interpretation as different ways of `cutting open' the loop in the diagram, and each term can be associated to a tree-diagram involving the (on-shell) momentum $\tilde\ell$; see \cref{fig:loopstotrees}. 

We discuss an explicit example of the above presciption for changing the representation of the loop integrand from $I_{qdr}$ to $I_{lin}$ in \cref{sec:YM}. \\

While non-standard, this novel representation of the loop integrand has several remarkable properties, as we will see below in detail: 
\begin{itemize}
 \item First and foremost, it facilitates a straightforward extension of tree-level results to loop-level. This is particularly evident in the construction of the BCJ numerators below, cf.~\cref{sec:algorithm}, but has also been implicitly used above in writing down the integrands \eqref{equ:loop-integrands} for Yang-Mills and gravity.
 \item Moreover, it manifests symmetry properties that are either obscured or absent in the usual representation. In particular, while the construction of local BCJ numerators on quadratic propagators faces serious obstacles for six external particles at one loop \cite{Mafra:2014gja, Berg:2016fui}, these difficulties are absent in this representation of the integrand; see also \cite{He:2017spx, He:2016mzd}. 
 \item As a last point, several checks on the amplitude are substantially simplified in this framework, and we exploit this in \cref{sec:YM}. Recall the discussion below \eqref{eq:loopintgeneral} about terms in the integrand that integrate to zero due to scaling with $\ell$, and consider also the form of $\mathfrak{I}$ in \eqref{eq:IlinQ}. The numerators are polynomials in $\ell\cdot k_i$ and $\ell\cdot \epsilon_i$, and the linear propagator factors always take the form $1/(2\ell\cdot K+K^2)$, where $K$ is a sum of external momenta. Therefore, only terms with at least one propagator factor for which $K^2\neq 0$ can contribute after the loop integration. In particular, any term for $n=2$ and $n=3$ must vanish upon integration because the Mandelstam variables are trivial.
\end{itemize}

Notice, however, that while it is straightforward to go from the quadratic propagators to the linear ones in the new representation, as shown in \eqref{eq:IlinQ}, the converse is not true. To our knowledge, there is no general algorithm to go in the opposite direction starting from contributions with linear propagators.

Refs.~\cite{Gomez:2017lhy,*Gomez:2017cpe} have explored the use of the scattering equations formalism to directly produce quadratic propagators at loop level, at the price of more elaborate CHY-type expressions. It would be interesting to know whether this idea can be efficiently applied to gauge theory and gravity, and whether it has a simple interpretation from the point of view of ambitwistor strings.


\section{One-loop BCJ numerators}\label{sec:nums}
The main advantage of the CHY representation \eqref{eq:loopint} and the resulting integrand $\mathfrak{I}$ lies in its simplicity: effectively, it has the complexity of an $(n+2)$-particle on-shell tree-amplitude. This close similarity makes it possible to easily extend tree-level results to loop level. 
In this section, we use this relation to derive an algorithm for all-multiplicity BCJ numerators, building on the corresponding work at tree level in \cite{Fu:2017uzt}.

\subsection{Derivation from ambitwistor strings}\label{sec:derivation}
Just as at tree level, the BCJ relations embed straightforwardly into the ambitwistor-string setting. For instance, 
\begin{equation}
 \sum_{j=1}^{n-1}\frac{\ell\cdot k_{12...j}}{\sigma_{12}...\sigma_{j+}\sigma_{+\,(j+1)}...\sigma_{n-}\sigma_{-1}}=0 \,,\qquad\text{mod }E_a^{(\text{nod})}\,,
\end{equation}
as shown in \cite{He:2016mzd}. 
Since the colour/kinematic duality consistency relations are naturally satisfied on the support of the scattering equations, we can follow the same main idea as at tree-level to derive the one-loop BCJ numerators.

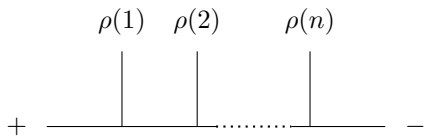
\begin{figure}[ht]
\begin{center}
 \begin{tikzpicture}[scale=1]
 \draw (0,0) -- (2.25,0) ;
 \draw[dotted, thick] (2.25,0) -- (3.25,0) ;
 \draw (3.25,0) -- (4.5,0);
 \draw (1,0) -- (1,1) ;
 \draw (2,0) -- (2,1) ;
 \draw (3.5,0) -- (3.5,1) ;
 \node at (-0.4,0) {$+$};
 \node at (4.9,0) {$-$};
 \node at (1,1.4) {$\rho(1)$};
 \node at (2,1.4) {$\rho(2)$};
 \node at (3.5,1.4) {$\rho(n)$};
\end{tikzpicture}
\end{center}
\caption{The one-loop BCJ master numerators $N\big(1,\rho(2,...,n-1),n\big)$ correspond to half-ladder diagrams with legs $+$ and $-$ at opposite endpoints. }
\label{fig:half-ladder-1loop}
\end{figure}

Our goal is therefore to expand both the Yang-Mills integrand $\mathfrak{I}_{\text{YM}}$ and the gravity integrand $\mathfrak{I}_{\text{grav}}$ in a DDM half-ladder basis, 
\begin{subequations} \label{equ:expansionloop}
\begin{align}
 \mathfrak{I}_{\text{YM}}&=\sum_{\rho\in S_n} c(+,\rho,-) \mathfrak{I}_{\text{YM}}(+,\rho,-) \,,
\label{equ:YMexpansionloop}
\\
 \mathfrak{I}_{\text{grav}}&=\sum_{\rho\in S_n} N(+,\rho,-) \mathfrak{I}_{\text{YM}}(+\rho,-) \,,\label{equ:gravexpansionloop}
\end{align}
\end{subequations}
where the factors $\mathfrak{I}_{\text{YM}}(+,\rho,-)$ are colour-ordered Yang-Mills integrands. Both the colour factors,
$$
c(+,\rho,-) = f^{a_+a_{\rho(1)}b_1} \, f^{b_1a_{\rho(2)}b_2} \cdots f^{b_{n-1}a_{\rho(n)}a_-}\, \delta^{a_+a_-} \,,
$$
and the kinematic numerators $N(+,\rho,-)$ are associated to half-ladder diagrams with legs $+$ and $-$ at opposite endpoints; see \cref{fig:half-ladder-1loop}.
This expansion naturally manifests the double-copy structure, now at the level of the one-loop integrand. As at tree level, the kinematic numerators $N(+,\rho,-)$ in \eqref{equ:gravexpansionloop} are known as the (one-loop) master numerators, and all numerators for other cubic diagrams follow from the requirement of colour-kinematic duality \eqref{equ:BCJtripletnum}.

Rephrasing the above expansion in terms of the ambitwistor string integrands $\mathcal{I}^{(1)}$, \cref{equ:expansionloop} is equivalent to
\begin{subequations}
\begin{align}
& \cI_{\text{SU}(N_c)}^{(1)} \; = \sum_{\rho\in S_{n}} 
\frac{ c(+,\rho,-) 
}{\sigma_{+\rho(1)}...\sigma_{\rho(n)-}\sigma_{-+}} \,,  \\
& \cI_{\text{NS}}^{(1)} \; =\; \sum_r \pf'(M_{\text{NS}}^r)
\; = \sum_{\rho\in S_{n}} 
\frac{ N(+,\rho,-)
}{\sigma_{+\rho(1)}...\sigma_{\rho(n)-}\sigma_{-+}} \,,
\qquad \text{mod }E_a^{(\text{nod})}\,.
\label{equ:expansionpfaffianloop}
\end{align}
\end{subequations}
Crucially, the extension of KLT orthogonality \cite{Cachazo:2012da,Cachazo:2013gna} to one loop \cite{He:2017spx, He:2016mzd} guarantees that this expansion of the Pfaffian into loop partial integrands exists. However, determining the coefficients $N$ is far from trivial, since \eqref{equ:expansionpfaffianloop} relies on the support of the scattering equations. Different strategies have been used at tree level and at one loop, including the use of cross-rational identities \cite{Bjerrum-Bohr:2016axv}, of a differential operator for residues \cite{Chen:2016fgi,*Chen:2017bug}, and of the forward limit for the CHY-type loop integrand in supersymmetric theories \cite{He:2017spx}.

The strategy we will implement here is to first use the definition of the (reduced) Pfaffian to expand the gravity loop integrand $\mathcal{I}^{(1)}_{\text{NS-NS}}$ into a sum over simpler Pfaffians with purely kinematic prefactors. In turn, the expansion of these Pfaffians into pure Yang-Mills integrands is well known, \cite{Fu:2017uzt}, and can be obtained by expanding recursively and fixing the expansion coefficients by gauge invariance.\footnote{At tree-level, the Pfaffians obtained in the expansion coincide with the single-trace contribution to the Einstein-Yang-Mills amplitude. We will see below that a suitable generalisation 
holds at one loop as well.}

As a starting point, recall the definition of the (CHY) Pfaffian as a sum over permutations\footnote{This representation of the CHY Pfaffian is easy to derive from the ambitwistor string, where all fermions in the vertex operators have to be contracted. With two fermions in each vertex operator, the sum over Wick contractions is equivalent to a sum over all contraction cycles. Using the equivalence between permutations and the product of cycles corresponding to the orbits of that permutation leads directly to \cref{equ:defpfaffian}. 
Alternatively, \cref{equ:defpfaffian} can be derived directly from writing the Pfaffian as a sum over perfect matchings: since $A_{ij}$, $B_{ij}$ and $C_{ij}$ all involve factors of $1/\sigma_{ij}$, perfect matchings exchanging these terms can be combined into cycles. 
See e.g. \cite{Lam:2016tlk} for an explicit discussion at tree level.}
\begin{equation}\label{equ:defpfaffian}
 \pf\big(M_\text{NS}^{r}\big)=\sum_{\rho\in S_{n+2}}(-1)^{\text{sgn}(\rho)}\frac{M_IM_J...M_K}{\sigma_I\sigma_J...\sigma_K}\,.
\end{equation}
Above, $I$, $J$ and $K$ denote closed cycles determined by the permutation, and the coefficients are given by
\begin{subequations}
\begin{align}
 &M_I=tr(I):=\tr(F_{i_1}...F_{i_{n_I}})\,, && \sigma_I=\sigma_{i_1i_2}\sigma_{i_2i_3}...\sigma_{i_{n_I}i_1}\,, && \text{for\;\;}n_I>1\,,\\
 &M_I=C_{ii}\,, && \sigma_I=1\,, && \text{for\;\;}n_I=1\,,
\end{align}
\end{subequations}
with $n_I=$ length$(I)$ and $F_i^{\mu\nu}=k_i^{\mu}\epsilon_i^{\nu}-k_i^{\nu}\epsilon_i^{\mu}$. For readability, we have used the notation $k_+=-k_-=\ell$ and $\epsilon_+=\epsilon^r\,,\,\epsilon_-=(\epsilon^r)^\dagger$. \Cref{equ:defpfaffian} immediately generalizes to the reduced Pfaffian $\mathcal{I}^{(1)}_{\text{NS}}$,
\begin{equation}
 \mathcal{I}^{(1)}_{\text{NS}}\equiv\sum_r\pf'\big(M_\text{NS}^{r}\big)=\sum_{\rho\in S_{n+2}^{+-}}(-1)^{\text{sgn}(\rho)}\frac{W_IM_J...M_K}{\sigma_I\sigma_J...\sigma_K}\,.
\end{equation}
In an important distinction from \cref{equ:defpfaffian}, $I$ now denotes the open cycle determined by the choice of removed rows and columns $\hat i, \hat j$ in the definition of the reduced Pfaffian \eqref{equ:Ikintree}, and $S_{n+2}^{+-}$ indicates that we are only summing over permutations exchanging $+$ and $-$ or keeping them fixed. A convenient choice for $\hat i$ and $\hat j$ are the rows and columns associated to the loop momentum $\ell$. With $\hat i=\sigma_+$ and $\hat j=\sigma_-$, the coefficient $W_I$ of the open cycle $I$ is given by
\begin{align}
 W_I=\sum_{r}\epsilon^{r}\cdot F_{i_1}...F_{i_{n_I}}\cdot (\epsilon^{r})^\dagger =  
\left\{
\begin{array}{rl} & \tr\Big(F_{i_1}...F_{i_{n_I}}\Big) \qquad \text{for}\quad n_I>0\,,\\ & D-2 \qquad\qquad\quad\;\;\, \text{for}\quad n_I=0\,,
    \end{array} \right.
\end{align}
where we have explicitly carried out the sum over the basis of polarisation vectors for the loop momenta, according to \eqref{equ:looppolsub}.

By decomposing the sum in $S_{n+2}^{+-}$, the NS-integrand $\mathcal{I}^{(1)}_{\text{NS}}$ thus becomes
 \begin{align}
 \mathcal{I}^{(1)}_{\text{NS}} &= \sum_I\left(\sum_{\rho\in S_I^{+-}}(-1)^{\sgn(\rho)}\frac{W_I}{\sigma_I}\right) \left(\sum_{\bar\rho\in S_{\bar I}}(-1)^{\sgn(\bar \rho)}\frac{M_J...M_K}{\sigma_J...\sigma_K}\right) \nonumber\\
 &=\sum_I\sum_{\rho\in S_I^{+-}}(-1)^{\sgn(\rho)}W_I\, \frac{\pf\big(M_{\bar I}\big)}{\sigma_I}
 \,.\label{equ:expansionpfaffianderiv}
 \end{align}
In the last line, we have used the definition \eqref{equ:defpfaffian} to rewrite the sum over permutations $S_{\bar I}$ as a Pfaffian over a matrix we denote by $M_{\bar I}$.\footnote{The Pfaffian $\pf\big(M_{\bar I}\big)/\sigma_I$ can be identified as the single-trace contribution to the Einstein-Yang-Mills (EYM) one-loop integrand $\mathcal{I}_{\text{EYM}}^{{(1)},\text{gluon}, n_{\tr}=1}$ with gluons running in the loop \cite{wip}. This directly extends the tree-level results of \cite{Fu:2017uzt}.} To relate this representation of the integrand $\mathcal{I}^{(1)}$ to the half-ladder form of \cref{equ:gravexpansionloop}, we need to express the Pfaffian $\pf\big(M_{\bar I}\big)/\sigma_I$ as a sum over $(n+2)$-particle Parke-Taylor factors with kinematic coefficients $\tilde{Y}_{\bar I}$. The full BCJ numerators are then given  by 
\begin{equation}\label{equ:schem_num_exp}
 N(+,\rho,-) = \sum_{I}W_I \,\tilde{Y}_{\bar I}\,.
\end{equation}
A procedure for obtaining these kinematic factors $\tilde{Y}_{\bar I}$ has been developed in \cite{Fu:2017uzt}, and we refer the interested reader to the original paper for details of the derivation. A point worth highlighting is that the factors $\tilde{Y}_{\bar I}$ are extracted from the Pfaffian $\pf\big(M_{\bar I}\big)/\sigma_I$ recursively, and thus depend on an (arbitrary) {\it reference ordering} RO. The set of BCJ numerators is therefore non-unique. The final amplitudes are of course invariant with respect to this choice. 

Since both the one-loop measure \eqref{equ:1-loop-measure} and the integrand \eqref{eq:MNS} structurally resemble $(n+2)$-particle tree-amplitudes, the algorithm for $\tilde{Y}_{\bar I}$ generalizes straightforwardly from tree-level to one loop, and we summarize the procedure in the next subsection \ref{sec:algorithm}.

\subsection{Algorithm for master numerators}\label{sec:algorithm}
We have seen in the last subsection how BCJ master numerators can be derived in the scattering-equations formalism at one loop.  Below, we summarise the resulting algorithm and discuss several examples. In section \ref{sec:YM}, we will see an example of how to construct the full loop integrand from the BCJ numerators. \\

Following \cref{sec:derivation}, the master numerators are given by half-ladder diagrams, with the loop momenta $\pm \ell$ forming the `handles' of the half-ladder; see \cref{fig:half-ladder-1loop}. They are therefore characterised unambiguously by a {\it colour ordering} 
\begin{equation}
 \text{CO}=(+\,a_1a_2\,...\,a_n\,-)\,,
\end{equation}
where $a_1$ is the left-most particle next to $+\ell$. 
As discussed in \cref{sec:derivation}, the kinematic factors $\tilde{Y}_{\bar I}$ are obtained recursively, which corresponds to 
\begin{enumerate}
 \item[(I)] Fixing a {\it reference ordering} (RO) to be used in the definition of all numerators, independently of their colour ordering. For example,
 \begin{equation}
  \text{RO} = (+\,1\,2\, ...\, n\, -)\,.
 \end{equation}
\end{enumerate}
To keep the algorithm and the formulae compact, it will be useful to introduce a notation for `particle $i$ is to the left of particle $j$ in a given colour or reference ordering'. We denote this by $i\triangleleft j$ (CO) and $i \dashv j$ (RO), respectively; see \cref{table:ord-rel}.\\
 
The master numerators \eqref{equ:schem_num_exp} (and in particular $\tilde{Y}_{\bar I}$)  naturally depend on both the colour ordering and the global choice of reference ordering. This dependence is best expressed in the form of {\it split orderings}, defined to encode the difference between the two orderings. They can be constructed as follows:
\begin{enumerate}
 \item[(II)] Decompose the set of all particles into a colour-ordered subset $I$ and its complement $\bar{I}$. For $\bar{I}$, then construct all disjoint decompositions into $R$ subsets $\alpha^{(r)}$ satisfying the following criteria:\footnote{For readability, we choose to identify the subsets by raised indices, and their elements by lowered ones.}
  \begin{enumerate}[(i)]
    \item $\bar I=\cup_{r=1}^R \alpha^{(r)}$.
    \item Each subset $\alpha^{(r)}$ respects the colour ordering:
    \begin{equation}
     \forall_{i<j}\;\;\alpha^{(r)}_i \triangleleft \alpha^{(r)}_j\,.
    \end{equation}
    \item The last elements of the subsets respect the reference ordering. Using again the notation $n_I=$ length$(I)$, and $n_r=$ length$(\alpha^{(r)})$, this can be compactly written as
    \begin{equation}
     \alpha^{(1)}_{n_1}\dashv ... \dashv \alpha^{(t)}_{n_R}\,.
    \end{equation}
    \item The last element of the subset is the smallest in the reference ordering. No ordering among the other elements is assumed; indeed the ordering is fixed by (ii).
    \begin{equation}
     \forall_i\;\; \alpha^{(r)}_{n_r}\dashv  \alpha^{(r)}_i\,.
    \end{equation}
  \end{enumerate}
 
  For any decomposition $\bar I=\cup_r \alpha^{(r)}$ satisfying these criteria, the {\it split ordering} (SO) is defined as 
  \begin{equation}
  \text{SO} = \left(+\,I\,\alpha^{(1)}\,...\,\alpha^{(R)}\,-\right)\,.
  \end{equation}
  As above, it will be convenient to introduce the notation $i\prec j$ to mean `particle $i$ is to the left of particle $j$ in the split ordering', c.f. \cref{table:ord-rel}.
  \begin{table}[ht]
  \begin{center}
  \begin{tabular}{|c|c|}
   \hline
   relation & `particle $i$ is to the left of particle $j$ in a given ...' \\ \hline
   $i\triangleleft j$  & colour ordering \\
   $i \dashv j$ & reference ordering \\
   $i\prec j$ & split ordering \\\hline
  \end{tabular}
  \caption{Notation for colour-ordering, reference-ordering and split-ordering relations.}
  \label{table:ord-rel}
  \end{center}
  \end{table}
 \item[(III)] Using these definitions, the master numerators are given by
 \begin{equation}\label{equ:numerators}
  N_{\text{RO}}(a_1\,a_2\,...\,a_n) = \sum_I (-1)^{n_I}\,W_I\left(\sum_{\text{SO}}\prod_{r}Y\left(\alpha^{(r)}\right)\right)\,,
 \end{equation}
 where the sum runs over split orderings $\text{SO} = \left(+\,I\,\alpha^{(1)}\,...\,\alpha^{(R)}\,-\right)$ and
 \begin{align}
  W_{I\neq \emptyset} &= \tr(I):=\tr(F_{i_1}\cdot...\cdot F_{i_{n_I}})\,,
\qquad W_{\emptyset} = D-2\,, \\
  Y\left(\alpha^{(r)}\right) &=\begin{cases} \epsilon_a \cdot Z_a & \alpha^{(r)}=\{a\}\\ \epsilon_{a_{n_r}}\cdot F_{a_{(n_r-1)}} \cdot...\cdot F_{a_1}\cdot Z_{a_1} & \alpha^{(r)}=\{a_1,...,a_{n_r}\}\,,\end{cases}
 \end{align}
where $F_i^{\mu\nu}=k_i^{\mu}\epsilon_i^{\nu}-k_i^{\nu}\epsilon_i^{\mu}$.
 The $Z_a$ are defined as the sum over momenta to the left of particle $a$ in {\it both} the colour ordering and the split ordering,
 \begin{equation}\label{equ:defZ}
  Z_a = \sum_i k_i \,\,\forall_{i\triangleleft a\text{ and }i\prec a}\,.
 \end{equation}
 This implies in particular that different terms in the sum over SO's in \cref{equ:numerators} may involve different sums over momenta for $Z_a$, since the split orderings differ.
\end{enumerate}
To highlight the relation to the schematic formula of the numerators as $N(+,\rho,-) = \sum_{I}W_I \,\tilde{Y}_{\bar I}$ given in \cref{equ:schem_num_exp}, note that \cref{equ:numerators} amounts to
\begin{equation}
 \tilde{Y}_{\bar I} = (-1)^{n_I} \left(\sum_{\text{SO}}\prod_{r}Y\left(\alpha^{(r)}\right)\right)\,.
\end{equation}
Relating this back to the expansion of the integrand $\mathcal{I}_{\text{NS}}^{(1)}$, the product over $Y\left(\alpha^{(r)}\right)$ can be understood intuitively as `breaking open' the closed cycles in the definition of the Pfaffian (see \cref{equ:expansionpfaffianderiv}), and the sum over split ordering ensures that only terms consistent with the colour ordering and the recursive expansion of the Pfaffian are kept.\\

Let us now take a closer look at the split ordering featuring so prominently in the algorithm. As pointed out at the beginning of this section, the split orderings are constructed to encode the difference between the two orderings the numerators depend on: the colour ordering CO and the reference ordering RO. This is best understood in a concrete example: consider the term coming from $\bar I = \{3,4\}$ in the four-point master numerators $N_{\text{RO}}(2134)$ and $N_{\text{RO}}(2143)$. With RO$=(+1234-)$, and hence $3\dashv 4$, the set Split$(\bar I)$ of possible decompositions $\bar I=\cup_r \alpha^{(r)}$ is
\begin{align}
 \text{Split}(\bar I) &= \begin{cases} \{\{3\},\{4\}\} & \text{for }3\triangleleft 4\,,\text{ so }N_{\text{RO}}(2134) \,, \\ 
                              \{\{3\},\{4\}\}\cup\{\{4,3\}\} & \text{for }4\triangleleft 3\,,\text{ so } N_{\text{RO}}(2143)\,,\end{cases}\\
 \text{and thus } \quad \text{SO} &=\begin{cases} (+2134-) & \text{for }3\triangleleft 4\,, \\ 
                              (+2134-),(+2143-) & \text{for }4\triangleleft 3\,.\end{cases} \label{eq:exampleorderings}
\end{align}
We see that if $\dashv$ and $\triangleleft$ agree when restricted to $\bar I$, there is a unique decomposition of $\bar I$ (with all subsets containing only a single element, $\forall_r\ n_r=1$). On the other hand, if $\dashv$ and $\triangleleft$ differ on $\bar I$, the decomposition contains subsets $\alpha^{(r)}$ with more than one element (e.g. $\{4,3\}$ in \cref{eq:exampleorderings}). In particular, this observation {\it only} depends on whether $\dashv$ and $\triangleleft$ describe different ordering relations on $\bar I$, without reference to $I$. Indeed, the above example is chosen such that CO $\neq$ RO for both numerators. \\

In other words, the set $\{\alpha^{(r)}\,|\,n_r>1\}$ is in one-to-one correspondence with the set of colour-ordered subsets of the external particles whose ordering differs from the chosen reference ordering RO. This will be a useful interpretation to keep in mind for the next subsection, where we calculate some of the master numerators for illustration.

\subsection{Examples}\label{sec:examples}
The easiest way to approach the rather abstract algorithm given in the last section is to consider a few examples. Along the way, we will make some useful observations concerning the form of the numerators. For simplicity, we take
$$
\text{RO} = (+1234-)
$$
for the remainder of this section.

\paragraph{The simplest example: $\mathbf{N_{\text{RO}}(1234)}$.} Keeping in mind the correspondence between $\alpha^{(r)}$ and colour-ordered, but not reference-ordered subsets of external particles, it is easy to see why $N_{\rm{RO}}(1234)$ constitutes the simplest case: since CO = RO, $\dashv$ and $\triangleleft$ agree when restricted to any $\bar I$, and thus the sum over split orderings only involves a single term for any $\bar I$. We see this explicitly in \cref{table:n1234contrib}.\\
\begin{table}[ht]
\begin{center}
\begin{tabular}{|l|l|c|c|l|}\hline
 $\bar I$ &  Split$(\bar I)$ & SO & $\tr(I)$ & numerator factor $W\left(\alpha^{(r)}\right)$ \\ \hline
 $\{1,2,3,4\} $ & $\{\{1\},\{2\},\{3\},\{4\}\}$ & $(+1234-)$ & $(D-2)$ & $\epsilon_1\cdot\ell\, \epsilon_2\cdot(\ell+k_1) \,\epsilon_3\cdot (\ell-k_4)\,\epsilon_4\cdot \ell$\\
 $\{1,2,3\}$ & $\{\{1\},\{2\},\{3\}\}$ & $(+4123-)$ & 0 & \;...\\
 \;\;\vdots & & & & \;\;\vdots\\
 $\{1,2\}$ & $\{\{1\},\{2\}\}$ & $(+3412-)$ & $\tr(34)$ & $\epsilon_1\cdot\ell\, \epsilon_2\cdot(\ell+k_1)$\\
 \;\;\vdots & & & & \;\;\vdots\\
  $\{1\}$ & $\{\{1\}\}$ & $(+2341-)$ & $\tr(234)$ & $\epsilon_1\cdot\ell$\\
 \;\;\vdots & & & & \;\;\vdots\\
 $\;\varnothing$ & $\{\}$ & $(+1234-)$& $\tr(1234)$ & \;\,1 \\ \hline
\end{tabular}
\end{center}
\caption{Contributions to the master numerator $N_{(1234)}(1234)$.}
\label{table:n1234contrib}
\end{table}\newline
While we have only listed one example term at any $n_I$ for illustration, it is easily checked that the decomposition into $\alpha^{(r)}$ is unique for the other terms as well. Due to CO = RO, only subsets of length one satisfy the criteria of (ii) being colour-ordered and (iv) the last element being the smallest in the reference ordering. 

We can already make a useful observation concerning the structure of the master numerators based on this example: due to $\tr(F_i)=0$, all terms from a decomposition with $n_I=1$ (and hence $n_{\bar I}=n-1$) vanish. This simplification occurs for any number of external particles, and for any colour ordering and reference ordering; we will therefore omit these terms from now on. 

Summing all contributions from \cref{table:n1234contrib}, $N_{\text{RO}}(1234)$ is given by
\begin{align}
 N_{\text{RO}}(1234) &= (D-2)\, \epsilon_1\cdot\ell\, \epsilon_2\cdot(\ell+k_1) \,\epsilon_3\cdot (\ell-k_4)\,\epsilon_4\cdot \ell\\
  &\qquad +\Big(\tr(12)\epsilon_3\cdot (\ell-k_4)\,\epsilon_4\cdot \ell+\tr(13)\epsilon_2\cdot(\ell+k_1)\,\epsilon_4\cdot\ell +\tr(14) \epsilon_2\cdot(\ell+k_1)\,\epsilon_3\cdot (\ell-k_4)  \nonumber\\
  &\qquad\qquad +tr(23)\epsilon_1\cdot\ell\,\epsilon_4\cdot\ell +\tr(24)\epsilon_1\cdot\ell\,\epsilon_3\cdot (\ell-k_4)+tr(34)\epsilon_1\cdot\ell\,\epsilon_2\cdot(\ell+k_1)\Big) \nonumber\\
  &\qquad - \Big(\tr(123)\epsilon_4\cdot\ell+\tr(124)\epsilon_3\cdot (\ell-k_4)+\tr(134)\epsilon_2\cdot(\ell+k_1) +\tr(234)\epsilon_1\cdot\ell\Big)\nonumber\\
  &\qquad +\tr(1234)\,. \nonumber
\end{align}

\paragraph{A more interesting example: $\mathbf{N_{\text{RO}}(1243)}$.} Having gained some intuition with the algorithm in the last example, let us now turn to a more interesting case with CO $\neq$ RO. For $ N_{(1234)}(1243)$, the algorithm of \cref{sec:algorithm} leads to the terms listed in \cref{table:n1243contrib}.\\
\begin{table}[ht] 
\begin{center}
\begin{tabular}{|l|l|c|c|l|}\hline
 $\bar I$ &  Split$(\bar I)$ & SO & $\tr(I)$ & numerator factor $W\left(\alpha^{(r)}\right)$ \\ \hline 
 $\{1,2,4,3\} $ & $\{\{1\},\{2\},\{3\},\{4\}\}$ & $(+1234-)$ & $(D-2)$ & $\epsilon_1\cdot\ell\, \epsilon_2\cdot(\ell+k_1) \,\epsilon_4\cdot (\ell-k_3)\,\epsilon_3\cdot (\ell-k_4)$\\\vspace{5pt}
  & $\{\{1\},\{2\},\{4,3\}\}$ & $(+1243-)$ & $(D-2)$ & $\epsilon_1\cdot\ell\, \epsilon_2\cdot(\ell+k_1) \,\epsilon_3\cdot F_4\cdot (\ell-k_3)$\\ 
 $\{1,2\}$ & $\{\{1\},\{2\}\}$ & $(+4312-)$ & $\tr(43)$ & $\epsilon_1\cdot\ell\, \epsilon_2\cdot(\ell+k_1)$\\
 $\{4,3\}$ & $\{\{3\},\{4\}\}$ & $(+1234-)$ & $\tr(12)$ & $\epsilon_4\cdot (\ell-k_3)\,\epsilon_3\cdot (\ell-k_4)$\\
 & $\{\{4,3\}\}$ & $(+1243-)$ & $\tr(12)$ & $\epsilon_3\cdot F_4\cdot (\ell-k_3)$\\
 \;\;\vdots & & & & \;\;\vdots\\
  $\{1\}$ & $\{\{1\}\}$ & $(+2431-)$ & $\tr(243)$ & $\epsilon_1\cdot\ell$\\
 \;\;\vdots & & & & \;\;\vdots\\
 $\;\varnothing$ &$\{\}$ & $(+1243-)$ & $\tr(1243)$ & \;\,1 \\ \hline
\end{tabular}
\end{center}
\caption{Contributions to the master numerator $N_{(1234)}(1243)$.}
\label{table:n1243contrib}
\end{table}\newline
Consistently with the intuition developed in the last section, we get terms from the decompositions $\{\{3\},\{4\}\}$ and $\{\{4,3\}\}$, since $\dashv$ and $\triangleleft$ describe different ordering relations on any $\bar I$ containing both particles 3 and 4. We notice, moreover, that terms from different split orderings of the same $\bar I$ combine nicely; for example the contribution from $\bar I = \{4,3\}$ is
\begin{align}\label{eq:SOcombination34}
 \tr(12)\left(\sum_{\text{SO}(43)}\prod_{r}W\left(\alpha^{(r)}\right)\right)
 &= \tr(12)\Big(\epsilon_4\cdot (\ell-k_3)\,\epsilon_3\cdot \ell - \epsilon_3\cdot\epsilon_4\, k_4\cdot (\ell-k_3)\Big)\,.
\end{align}
The first term on the right-hand side actually looks familiar: we have already encountered it in $N_{(1243)}(1243)$ (with RO = CO).  A similar calculation for $\bar I = \{1,2,4,3\}$ shows that the two terms from SO = $(+1234-)$ and SO = $(+1243-)$ combine to the $N_{(1243)}(1243)$-contribution and terms proportional to $\epsilon_3\cdot \epsilon_4$. Combining all terms, the numerator is given by
\begin{align}
  N_{(1234)}(1243) &= N_{(1243)}(1243)-(D-2)\,\epsilon_1\cdot\ell\, \epsilon_2\cdot(\ell+k_1) \, \epsilon_3\cdot\epsilon_4\,k_4\cdot(\ell-k_3)\\
   &\hspace{70pt}-\tr(12)\,\epsilon_3\cdot\epsilon_4\,k_4\cdot(\ell-k_3) \,. \nonumber
\end{align}

\paragraph{Remark.} In fact, the observations from the examples above can be extended to any colour ordering and any number of external particles:
\begin{itemize}
 \item All terms with $n_I=1$ in the sum over split orderings vanish due to $\tr(F_I)=0$.
 \item Given a reference ordering RO and a colour ordering CO, the master numerators take the form
 \begin{equation}
  N_{\text{RO}}(\text{CO})=N_{\text{CO}}(\text{CO})-\Delta\,,\label{eq:numstructure}
 \end{equation}
 where all terms in $\Delta$ are proportional to $\epsilon_i\cdot\epsilon_j$ with $i\triangleleft j$ but $j\dashv i$.\footnote{Or explicitly, $i$ is to the left of $j$ in the colour ordering, but to the right of $j$ in the reference ordering.}
\end{itemize}
{\it Proof.} \Cref{eq:numstructure} is easily proven directly from the algorithm as follows. First note that it suffices to consider a single set $\bar I$ on which $\triangleleft$ and $\dashv$ differ, since the simplifications work term by term in the expansion $I\cup\bar I$. Moreover, although there is always a contribution coming from the decompositions where all subsets $\alpha^{(r)}$ have length one, this term does not directly coincide with $N_{\text{CO}}(\text{CO})$ due to the different split orderings. To be precise, $\Delta Z:=Z_{a}^{\text{CO}}-Z_{a}^{\text{RO}}=\sum_{i|i\triangleleft a,\, a \prec i}k_i$ is given by the sum over momenta respecting the colour ordering, $i\triangleleft a$, but violating the reference ordering, $a \prec i$. However, we will see that this is exactly compensated by the split orderings with $n_r>1$.

By construction, contributions from $n_r>1$ are of the form $\epsilon_{a_{n_r}}\cdot F_{(a_{n_r}-1)}\cdot...\cdot Z_{a_1}$ with $a_i$ satisfing $a_i\triangleleft a_1,\, a_1 \prec a_i$ for $i>1$. We further observe that $Z_{a_1}^{\text{CO}}= Z_{a_1}^{\text{RO}}$. Upon expanding $F_i^{\mu\nu} =k_i^{[\mu}\epsilon_i^{\nu]}$, the terms contracting {\it only} polarisation vectors to momenta thus add precisely the factors $\epsilon\cdot\Delta Z$ required to match to the numerators $N_{\text{CO}}(\text{CO})$. Collecting all terms containing the contractions {\it among} polarisation vectors as $\Delta$ then completes the proof of \cref{eq:numstructure}.

\paragraph{Corollary: All-plus and one-minus.}
The above remark leads to an immediate corollary for the all-plus helicities and one-minus helicities master numerators in four spacetime dimensions. For these amplitudes, the reference spinors in the polarisation vectors can be chosen such that $\epsilon_i\cdot\epsilon_j=0$ $\forall_{i,j}$; see \cref{equ:polarisation4d} below. Using \cref{eq:numstructure}, the numerators simplify to 
\begin{equation}
  N_{\text{RO}}(\text{CO})=N_{\text{CO}}(\text{CO})\,, \qquad \forall \,\text{RO} \,,
 \end{equation}
that is, the numerators are independent of the choice of reference ordering. This vastly simplifies the structure of the amplitudes and facilitates the matching to known results, which we will verify in the next section.

\section{Tests of our Yang-Mills formulae}\label{sec:YM}

In this section, we provide several explicit tests of our gauge theory results. The first part is dedicated to tests of our four-gluon scattering formula, whereas the second part is dedicated to the simplest four-dimensional helicity configurations: all-plus and one-minus (and analogously, all-minus and one-plus).

\subsection{Tests at four points} \label{subsec:YM4pt}

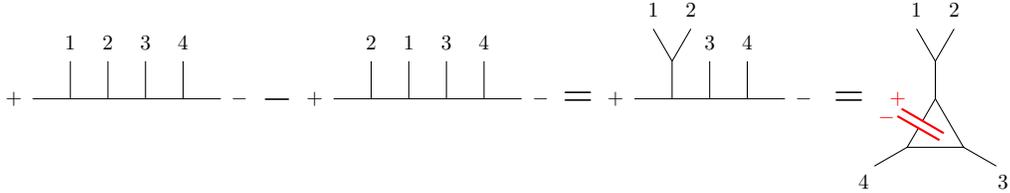
\begin{figure}[ht]
\begin{center}
 \begin{tikzpicture}[scale=0.5]
 \draw (0,0) -- (5,0) ;
 \draw (1,0) -- (1,1);
 \draw (2,0) -- (2,1);
 \draw (3,0) -- (3,1);
 \draw (4,0) -- (4,1);
 \node at (-0.5,0) {\scalebox{0.8}{$+$}};
 \node at (5.5,0) {\scalebox{0.8}{$-$}};
 \node at (1,1.5) {\scalebox{0.8}{$1$}};
 \node at (2,1.5) {\scalebox{0.8}{$2$}};
 \node at (3,1.5) {\scalebox{0.8}{$3$}};
 \node at (4,1.5) {\scalebox{0.8}{$4$}};
 \node at (6.5,0) {\scalebox{1.5}{$\mathbf{-}$}};
 \draw (8,0) -- (13,0);
 \draw (9,0) -- (9,1);
 \draw (10,0) -- (10,1);
 \draw (11,0) -- (11,1);
 \draw (12,0) -- (12,1);
 \node at (7.5,0) {\scalebox{0.8}{$+$}};
 \node at (13.5,0) {\scalebox{0.8}{$-$}};
 \node at (9,1.5) {\scalebox{0.8}{$2$}};
 \node at (10,1.5) {\scalebox{0.8}{$1$}};
 \node at (11,1.5) {\scalebox{0.8}{$3$}};
 \node at (12,1.5) {\scalebox{0.8}{$4$}};
 \node at (14.5,0)  {\scalebox{1.5}{$\mathbf{=}$}};
 \draw (16,0) -- (20,0);
 \draw (17,0) -- (17,1) -- (16.5,1.86);
 \draw (17,1) -- (17.5,1.86);
 \draw (18,0) -- (18,1);
 \draw (19,0) -- (19,1);
 \node at (15.5,0) {\scalebox{0.8}{$+$}};
 \node at (20.5,0) {\scalebox{0.8}{$-$}};
 \node at (16.5,2.36) {\scalebox{0.8}{$1$}};
 \node at (17.5,2.36) {\scalebox{0.8}{$2$}};
 \node at (18,1.5) {\scalebox{0.8}{$3$}};
 \node at (19,1.5) {\scalebox{0.8}{$4$}};
 \node at (21.7,0)  {\scalebox{1.5}{$\mathbf{=}$}};
 \draw (23.662,-0.585) -- (24,0) -- (24,1) -- (23.5,1.86);
 \draw (24,1) -- (24.5,1.86);
 \draw (24,0) -- (24.75,-1.29) -- (23.25,-1.29) -- (23.55,-0.78);
 \draw[red,thick] (23.103,-0.265) -- (24.223,-0.915);
 \draw[red,thick] (22.99,-0.46) -- (24.11,-1.1);
 \draw (24.75,-1.29) -- (25.62,-1.79);
 \draw (23.25,-1.29) -- (22.38,-1.79);
 \node[red,thick] at (23,0) {\scalebox{0.8}{$+$}};
 \node[red,thick] at (22.7,-0.5) {\scalebox{0.8}{$-$}};
 \node at (23.5,2.36) {\scalebox{0.8}{$1$}};
 \node at (24.5,2.36) {\scalebox{0.8}{$2$}};
 \node at (25.8,-2.2) {\scalebox{0.8}{$3$}};
 \node at (22.1,-2.2) {\scalebox{0.8}{$4$}};
\end{tikzpicture}
\end{center}\vspace{-20pt}
\caption{Graphic depiction of the Jacobi identity giving rise to the triangle numerator in \cref{equ:jac_triangle}.
}
\label{fig:trianglejacobis}
\end{figure}

The four-point colour-ordered one-loop integrand for gluons is given by a combination of box, triangle and bubble diagram contributions, all defined with respect to the numerators $N$ defined in the previous section. It reads\footnote{The explicit average in the sign of $\ell$ in \eqref{eq:IYM1234} is not essential, but leads to convenient cancellations, simplifying the tests to be described below. For instance, we find that the sign average is not needed to check the cyclic relation ${\mathfrak{I}}_{\text{YM}}(1234)={\mathfrak{I}}_{\text{YM}}(2341)$, but it enforces directly the reflection relation ${\mathfrak{I}}_{\text{YM}}(1234)={\mathfrak{I}}_{\text{YM}}(4321)$.}
\begin{align}
{\mathfrak{I}}_{\text{YM}}(1234) =& \, \frac{1}{2}\,\Big[\, \mathfrak{I}^{\text{box}}_{\text{YM}}(1234) + \left(\mathfrak{I}^{\text{tri}}_{\text{YM}}(1234)+\mathfrak{I}^{\text{tri}}_{\text{YM}}(2341)+\mathfrak{I}^{\text{tri}}_{\text{YM}}(3412)+\mathfrak{I}^{\text{tri}}_{\text{YM}}(4123)\right) \nonumber \\ &
\qquad +\left(\mathfrak{I}^{\text{bub}}_{\text{YM}}(1234)+\mathfrak{I}^{\text{bub}}_{\text{YM}}(4123) \right)
\; + (\ell\to-\ell) \,\Big]
\ ,
\label{eq:IYM1234}
\end{align}
and
\begin{align}
\mathfrak{I}^{\text{box}}_{\text{YM}}(1234) =& \  \frac{N(1234;\ell)}{(2\ell\cdot k_1)(2\ell\cdot (k_1+k_2)+2k_1\cdot k_2)(-2\ell\cdot k_4)} + \text{cyc}(1234)\ ,
\nonumber \\
\mathfrak{I}^{\text{tri}}_{\text{YM}}(1234) =& \ \frac{1}{2k_1\cdot k_2} \Bigg(
\frac{N([1,2]34;\ell)}{(2\ell\cdot (k_1+k_2)+2k_1\cdot k_2)(-2\ell\cdot k_4)}
\nonumber \\
& \qquad\qquad +   \frac{N(34[1,2];\ell)}{(2\ell\cdot k_3)(2\ell\cdot (k_3+k_4)+2k_3\cdot k_4)}  + \frac{N(4[1,2]3;\ell)}{(2\ell\cdot k_4)(-2\ell\cdot k_3)}
 \Bigg) \ ,\nonumber \\
\mathfrak{I}^{\text{bub}}_{\text{YM}}(1234) =& \ \frac{1}{(2k_1\cdot k_2)^2} 
\Bigg( \frac{N([1,2][3,4];\ell)}{2\ell\cdot (k_1+k_2)+2k_1\cdot k_2} + \frac{N([3,4][1,2];\ell)}{2\ell\cdot (k_3+k_4)+2k_3\cdot k_4}  \Bigg)\nonumber\ ,
\end{align}
where $\text{cyc}(1234)$ denotes the three remaining cyclic permutations. The full (colour-dressed) integrand, including the double-trace contributions, is obtained from \eqref{equ:YMexpansionloop}, where  colour-ordered integrands ${\mathfrak{I}}_{\text{YM}}(+\rho-)$ are now denoted as ${\mathfrak{I}}_{\text{YM}}(\rho)$ for brevity. In the expressions above, we introduced two pieces of notation. Firstly, we wrote explicitly the loop momentum in the numerators, $N(...;\ell)$, which will be helpful later for comparison with a different representation. Secondly, we incorporated the BCJ Jacobi relations by defining the triangle and bubble numerators in terms of the box numerators constructed in the previous section, e.g.
\begin{subequations}
\begin{align}
N([1,2]34;\ell) =& \ N(1234;\ell) - N(2134;\ell) \ , \label{equ:jac_triangle}\\
N([1,2][3,4];\ell) =& \ N(1234;\ell) - N(2134;\ell) - N(1243;\ell) + N(2143;\ell) \ ,
\end{align}
\end{subequations}
and so on; see fig.~\ref{fig:trianglejacobis} for a graphic depiction of the first line. The box numerators are simply the half-ladder numerators at four points.

We performed three basic numerical tests of our formulae, using random kinematic points. The first test is gauge invariance. Working with any particular reference ordering (RO) to define the numerators, we find numerically that a gauge transformation changes the integrand but not the amplitude:
\begin{equation}
{\mathfrak{I}}_{\text{YM}}(1234) \Big|_{\epsilon_i\to\epsilon_i+\alpha _ik_i}  \simeq {\mathfrak{I}}_{\text{YM}}(1234)\ .
\end{equation}
Explicitly, we check that the difference between the left-hand and right-hand sides in the expression above either scales for a rescaling of the loop momentum or is a linear combination of terms that scale; equivalently, that there is no propagator factor of the type $2\ell\cdot K+K^2$ with non-vanishing $K^2$, since this would ruin the scaling. For the present test, we find simply
\begin{equation}
{\mathfrak{I}}_{\text{YM}}(1234) \Big|_{\epsilon_i\to\epsilon_i+\alpha _ik_i}  -\, {\mathfrak{I}}_{\text{YM}}(1234) \; \stackrel{\ell\to\lambda\ell}{\longrightarrow} \; {\mathcal O}(\lambda^0) \ ,
\end{equation}
but in other tests below we may find (finite) linear combinations of terms ${\mathcal O}(\lambda^m)$ for integer $m$. According to the argument around \eqref{eq:Delta0}, this ensures that the two loop integrands are equivalent. The simplicity of this important test is striking, when compared with the more elaborate procedure required to verify gauge invariance in a Feynman-type representation of the integrand.

The second test we performed is the verification that we obtain an equivalent loop integrand ($\simeq$) for any choice of reference ordering in the definition of the numerators. For instance,\footnote{We denote the reference ordering $(+i_1i_2i_3i_4-)$ simply as $(i_1i_2i_3i_4)$ for brevity.}
\begin{equation}
{\mathfrak{I}}_{\text{YM}}(1234) \Big|_{\RO=(1234)}  \simeq {\mathfrak{I}}_{\text{YM}}(1234) \Big|_{\RO=(4231)} \ .
\end{equation}
More generally, we can use any linear combination
\begin{equation}\label{N_CO_ICO}
N(\text{CO};\ell) = \sum_{\RO \in S_4} \left( c^{(\RO)} \ N_{\RO}(\text{CO};\ell) \ + \ \tilde c^{(\RO)} \ N_{\RO}(\text{ICO};-\ell) \right) \ , \qquad \forall \text{CO}\in S_4 \ ,
\end{equation}
as long as the constant coefficients satisfy
\begin{equation}
\sum_{\RO \in S_4} \left( c^{(\RO)} + \tilde c^{(\RO)} \right) =1\ .
\end{equation}
We denote by $\text{ICO}$ the inverted colour ordering; for instance, $\text{CO}=(1234) \Rightarrow \text{ICO}=(4321)$. The inclusion of these terms into the numerators in \cref{N_CO_ICO} is allowed because of the reflection property of the propagators in half-ladder diagrams (which requires momentum conservation), e.g.
$$ 
(2\ell\cdot k_4)(2\ell\cdot (k_4+k_3)+2k_4\cdot k_3)(-2\ell\cdot k_1) \Big|_{\,\ell\to -\ell}
=
(2\ell\cdot k_1)(2\ell\cdot (k_1+k_2)+2k_1\cdot k_2)(-2\ell\cdot k_4) \,.
$$
Let us give an explicit example. If we define the numerators by choosing the linear combination
\begin{equation}
N(\text{CO};\ell) = (1-c) \ N_{(2413)}(\text{CO};\ell) + c \ N_{(4312)}(\text{ICO};-\ell) \ , \qquad \forall \text{CO}\in S_4 \ ,
\end{equation}
then the terms proportional to the coefficient $c$ in the complete integrant ${\mathfrak{I}}_{\text{YM}}(1234)$ vanish after loop integration, due to their scaling with the loop momentum, as discussed above. A natural choice of numerators is to symmetrise over all reference orderings, as we discuss in appendix \ref{sec:mastersym}. In this case, the various colour orderings are obtained from each other by just relabelling the particles.

For the third test, we match our loop integrand (in any of the equivalent forms described above) with a known form of the four-point integrand. That is, while we have verified the consistency of our expressions with respect to gauge transformations and to the choice of reference ordering that defines the numerators, we are still to show explicitly that it matches the correct amplitude. In order to do this, we choose a conventional BCJ representation of the integrand -- that is, one with Feynman-type quadratic propagators -- obtained in \cite{Bern:2013yya}. In particular, we define as
$$
\hat{N}(1234;\ell;\ell^2)
$$
the box BCJ numerator of equation (3.5) in that paper. Notice that, in this conventional BCJ representation, the analogous numerator for any other ordering of the particles is given by a simple relabelling. Moreover, the numerators $\hat{N}$ depend on $\ell^2$ (unlike ours, which depend on $\ell$ only through $\ell\cdot k_i$ and $\ell\cdot \epsilon_i$) and we denote this dependence explicitly in its arguments. Now, in order to compare to our numerators, we must change from the Feynman-type representation $I_{qdr}$ of the integrand used in \cite{Bern:2013yya} into the representation $I_{lin}$ we use. We can follow the procedure reviewed around \eqref{eq:partialfrac} to obtain
\begin{align}
\hat{\mathfrak{I}}_{\text{YM}}(1234) =& \, \frac{1}{2}\,\Big[\, \hat{\mathfrak{I}}^{\text{box}}_{\text{YM}}(1234) + \left(\hat{\mathfrak{I}}^{\text{tri}}_{\text{YM}}(1234)+\hat{\mathfrak{I}}^{\text{tri}}_{\text{YM}}(2341)+\hat{\mathfrak{I}}^{\text{tri}}_{\text{YM}}(3412)+\hat{\mathfrak{I}}^{\text{tri}}_{\text{YM}}(4123)\right) \nonumber \\ &
\qquad +\left(\hat{\mathfrak{I}}^{\text{bub}}_{\text{YM}}(1234)+\hat{\mathfrak{I}}^{\text{bub}}_{\text{YM}}(4123) \right)
\; + (\ell\to-\ell) \,\Big]
\ ,
\label{eq:IYM1234zvi}
\end{align}
and
\begin{align}
\hat{\mathfrak{I}}^{\text{box}}_{\text{YM}}(1234) =& \  \frac{\hat{N}(1234;\ell;0)}{(2\ell\cdot k_1)(2\ell\cdot (k_1+k_2)+2k_1\cdot k_2)(-2\ell\cdot k_4)} + \frac{\hat{N}(1234;\ell-k_1;-2\ell\cdot k_1)}{(2\ell\cdot k_2)(2\ell\cdot (k_2+k_3)+2k_2\cdot k_3)(-2\ell\cdot k_1)}
\nonumber \\
& \hspace{-1.7cm}+ \ \frac{\hat{N}(1234;\ell-k_1-k_2;-2\ell\cdot (k_1+k_2)+2k_1\cdot k_2)}{(2\ell\cdot k_3)(2\ell\cdot (k_3+k_4)+2k_3\cdot k_4)(-2\ell\cdot k_2)} + \frac{\hat{N}(1234;\ell+k_4;2\ell\cdot k_4)}{(2\ell\cdot k_4)(2\ell\cdot (k_4+k_1)+2k_4\cdot k_1)(-2\ell\cdot k_3)}
\ , \nonumber \\
\hat{\mathfrak{I}}^{\text{tri}}_{\text{YM}}(1234) =& \ \frac{1}{2k_1\cdot k_2} \Bigg(
\frac{\hat{N}([1,2]34;\ell;0)}{(2\ell\cdot (k_1+k_2)+2k_1\cdot k_2)(-2\ell\cdot k_4)}
\nonumber \\
& \quad +   \frac{\hat{N}([1,2]34;\ell-k_1-k_2;-2\ell\cdot (k_1+k_2)+2k_1\cdot k_2)}{(2\ell\cdot k_3)(2\ell\cdot (k_3+k_4)+2k_3\cdot k_4)}  + \frac{\hat{N}([1,2]34;\ell+k_4;2\ell\cdot k_4)}{(2\ell\cdot k_4)(-2\ell\cdot k_3)}
 \Bigg) \ ,\nonumber \\
\hat{\mathfrak{I}}^{\text{bub}}_{\text{YM}}(1234) =& \ \frac{1}{(2k_1\cdot k_2)^2} 
\Bigg( \frac{\hat{N}([1,2][3,4];\ell;0)}{2\ell\cdot (k_1+k_2)+2k_1\cdot k_2} 
\nonumber \\
& \quad + \frac{\hat{N}([1,2][3,4];\ell-k_1-k_2;-2\ell\cdot (k_1+k_2)+2k_1\cdot k_2)}{2\ell\cdot (k_3+k_4)+2k_3\cdot k_4}  \Bigg)\nonumber\ .
\end{align}
The choice of the last argument in the numerators $\hat{N}$ is associated with the shifts $\ell^2\to\ell^2+\eta^2$ for each term, as explained below \eqref{eq:partialfrac}.
Finally, we can verify numerically that
\begin{equation}
\hat{{\mathfrak{I}}}_{\text{YM}}(1234) \simeq {\mathfrak{I}}_{\text{YM}}(1234)\ ,
\end{equation}
where our integrand ${\mathfrak{I}}_{\text{YM}}$ can be given by any reference ordering of the numerators $N$, or by any valid linear cobination of these as discussed above.

We conclude that we have the correct four-particle one-loop integrand.

For completeness, we performed one more check: the verification of the one-loop BCJ relations. For the representation of the loop integrand that we use, these relations were given in \cite{He:2016mzd}. We find that the relations (5) and (6) in that paper are still valid, but only up to terms that integrate to zero due to scaling. This subtlety does not occur for the supersymmetric case studied in \cite{He:2016mzd}, but was anticipated in previous work on the one-loop BCJ relations \cite{Boels:2011tp, *Boels:2011mn}. Likewise, we find that the partial integrands defined in \cite{He:2016mzd} are now gauge invariant only up to terms that integrate to zero.

\subsection{Tests in four dimensions: all-plus and one-minus amplitudes} \label{subsec:YMallplus}

The simplest four-dimensional amplitudes are those for which all the particles have the same helicity. In a helicity basis for the polarisations, for which we take a spinor-helicity representation,
\begin{equation}
\epsilon_i^{(+)}= \frac{|\eta\rangle [i|}{\langle i \eta \rangle} \ , \qquad 
\epsilon_i^{(-)}= \frac{|i\rangle [\eta|}{[ i \eta ]} \ ,\label{equ:polarisation4d}
\end{equation}
we have $\epsilon_i^\pm \cdot \epsilon_j^\pm=0$. As we saw in \eqref{eq:numstructure}, the definition of the numerators is independent of the choice of reference orderings when $\epsilon_i \cdot \epsilon_j=0$. In the case where all the particles have the same helicity, we have verified up to 15 points that
\begin{equation}
N(1^+2^+\cdots n^+;\ell) = 2\ \prod_{i=1}^n \frac{1}{\langle \eta i\rangle^2} \ X(\ell+k_1+\cdots + k_{i-1},k_i) \ ,
\end{equation}
where the factor 2 comes from $D-2$, and we made use of the object
\begin{equation}
X(K,K') \equiv -\langle \eta | K K'  | \eta \rangle = -X(K',K) \ ,
\end{equation}
with the momenta $K$ and $K'$ possibly off-shell. These numerators are precisely the BCJ numerators for all-plus amplitudes found in \cite{Boels:2013bi}. The all-plus one-loop amplitudes belong to the self-dual sector of Yang-Mills theory, where $X(K,K') $ plays the role of vertex in diagrams. As first described in \cite{Monteiro:2011pc}, the colour-kinematics duality in the self-dual sector is a consequence of the (Schouten) identity
\begin{equation}
X(K_a,K_b) \ X(K_c,K_d) + X(K_b,K_c) \ X(K_a,K_d) + X(K_c,K_a) \ X(K_b,K_d) =0 \ ,
\end{equation}
which leads to the Jacobi relations among numerators.

The above result can be extended to the case when all but one particles have the same helicity. Let us take particle 1 to have negative helicity, while the others have positive helicity. We can still obtain $\epsilon_i \cdot \epsilon_j=0$ for any particles $i$ and $j$, if we choose $|\eta\rangle=|1\rangle$. Then our BCJ-numerator algorithm leads to the one-minus BCJ numerators also described in \cite{Boels:2013bi}, for instance,
\begin{equation}
N(1^-2^+\cdots n^+;\ell) = 2 \ \frac{1}{[ \eta 1]^2}\ \bar{X}(\ell,k_1) \prod_{i=2}^n \frac{1}{\langle \eta i\rangle^2} \ X(\ell+k_1+\cdots + k_{i-1},k_i) \ ,
\end{equation}
with $\bar{X}(K,K')\equiv -[ \eta | K K'  | \eta ]$, which we have also checked up to 15 points.

To conclude, the only non-supersymmetric families of amplitudes for which an $n$-point BCJ form was known explicitly -- the all-plus and one-minus sectors \cite{Boels:2013bi} -- are precisely reproduced by the general numerators that we introduce in this paper.

\subsection{Comment on representations of the loop integrand}

The all-plus and one-minus numerators we have just described are special in that they are valid both in our representation of the loop integrand and in a conventional representation with quadratic propagators, where they were originally found in \cite{Boels:2013bi}. To appreciate this, consider these two types of representation at four points, in \eqref{eq:IYM1234} and \eqref{eq:IYM1234zvi}, respectively. Our numerators obey
\begin{equation}
N(1234;\ell-k_1) = N(2341;\ell) \quad \text{for all-plus or one-minus helicities,}
\end{equation}
where the choice of reference ordering is irrelevant. However, in general, the expression above is only valid for certain distinct reference orderings, such as
\begin{equation}
N_{(1234)}(1234;\ell-k_1) = N_{(2341)}(2341;\ell) \ .
\end{equation}
For this reason, it is in general not easy to relate our numerators to a set of numerators in a conventional representation of the integrand with quadratic propagators. We expect that, in a conventional representation, BCJ numerators should in general depend on $\ell^2$, as in the four-point numerators $\hat{N}$ of \cite{Bern:2013yya} used above to test our loop integrand. It would be interesting to find a modification of our numerators that applies to a conventional integrand.

\subsection{Comment on unitarity cuts}

Unitarity based techniques are a powerful tool to deal with loop-level amplitudes \cite{Bern:1994zx,*Bern:1994cg,*Britto:2004nc,*Ellis:2007qk}. A natural question is whether our formulae are well suited to the application of these techniques, as an alternative to direct loop integration. Although this question is beyond the scope of this paper, we can make a simple remark.

Consider the four-point integrand given above in \eqref{subsec:YM4pt}. There exist four terms in the box contribution, whose propagators are cyclically related; the same does not apply in general to the numerators, as they depend on the choice of reference ordering. However, this complication is ultimately spurious. A four-dimensional maximal (box) cut is given by, for instance,\footnote{A non-vanishing result requires that there are two particles external of each helicity.}
\begin{itemize}
\item  either $\,N_{\text{RO}}(1234;\ell_a)\,$, independently of RO, with cut loop momentum solutions $\ell_a$, $a=1,2$,
\begin{equation}
(\ell_a)^2=2\,\ell_a\cdot k_1=2\,\ell_a\cdot (k_1+k_2)+2\, k_1\cdot k_2 = -2\,\ell_a\cdot k_4 =0 \,,
\end{equation}
\item  or $\,N_{\text{RO'}}(2341;\ell'_b)\,$, independently of RO', with cut loop momentum solutions $\ell'_b$, $b=1,2$,
\begin{equation}
(\ell'_b)^2=2\,\ell'_b\cdot k_2=2\,\ell'_b\cdot (k_2+k_3)+2\, k_2\cdot k_3 = -2\,\ell'_b\cdot k_1 =0 \,.
\end{equation}
\end{itemize}
The result is equivalent. This shows that the potentially unattractive features of the number of terms in our formulae (e.g.~four terms for each box diagram) and of the choice of reference ordering are not serious obstructions. The procedure above does not require a sum over internal polarisation states, and its connection to tree-level formulae is manifest from the algorithm for the numerators. We plan to explore the unitarity properties of our formulae elsewhere.

\section{Double copy to gravity} \label{sec:doublecopy}

The gravity loop integrand is obtained from the Yang-Mills case via the double-copy prescription, by squaring the numerators. The straightforward double copy gives the NS-NS gravity case. For four external particles, we have
\begin{align}
{\mathfrak{I}}_{\text{NS-NS},4} =& \, \frac{1}{2}\,\Big[\, \sum_{\text{CO}\in S_4} \left( \frac{1}{8}\ \mathfrak{I}^{\text{box}}_{\text{NS-NS}}(\text{CO}) + \frac{1}{4}\ \mathfrak{I}^{\text{tri}}_{\text{NS-NS}}(\text{CO})+ \frac{1}{16}\ \mathfrak{I}^{\text{bub}}_{\text{NS-NS}}(\text{CO}) \right)
\; + (\ell\to-\ell) \,\Big]
\ ,
\end{align}
where the box, triangle and bubble contributions are given from their Yang-Mills counterparts as
$$
\left\{ \ \mathfrak{I}^{\text{box}}_{\text{YM}} \ ,\ \mathfrak{I}^{\text{tri}}_{\text{YM}} \ ,\ \mathfrak{I}^{\text{bub}}_{\text{YM}}  \ \right\} \Big|_{\ N(\{\epsilon_i,k_i\};\ell)\ \to \ N(\{\epsilon_i,k_i\};\ell) \ N(\{\tilde\epsilon_i,k_i\};\ell)}
\ = \ \left\{ \ \mathfrak{I}^{\text{box}}_{\text{NS-NS}} \ ,\ \mathfrak{I}^{\text{tri}}_{\text{NS-NS}} \ ,\ \mathfrak{I}^{\text{bub}}_{\text{NS-NS}}  \ \right\}  \ .
$$
This gives the amplitude for the scattering of external states with factorisable polarisation tensors, $\varepsilon_{\mu\nu}=\epsilon_\mu\tilde\epsilon_\nu$. For more general states (where the polarisation tensors are not factorisable but can be given by linear combinations of factorisable contributions), we take appropriate linear combinations of the numerators for the factorisable states.

We have performed the analogous numerical tests as for the Yang-Mills four-point amplitude:
\begin{itemize}
\item invariance with respect to gauge transformations, now of both $\epsilon$ and $\tilde\epsilon$;
\item equivalence of different choices of reference ordering (or valid linear combinations of such choices) in the definition of the numerators, independently for left and right numerator factors in the double copy;
\item matching to a known form of the integrand, \cite{Bern:2013yya}, namely the gravity integrand obtained from the double copy of  \eqref{eq:IYM1234zvi}. 
\end{itemize}
In all these cases, the principle of the test is the same: the loop integrand ${\mathfrak{I}}_{\text{NS-NS},4}$ is only defined up to terms that integrate to zero according to their scaling with the loop momentum.
Moreover, the higher-point tests for the four-dimensional all-plus and one-minus helicity amplitudes apply also to the gravity case.

We have so far discussed the one-loop integrand for NS-NS gravity, where not only the graviton but also the dilaton and B-field run in the loop. Let us now present the one-loop integrand for pure gravity, with only gravitons in the loop. This can be performed by an explicit subtraction of the unwanted degrees of freedom. Recall from \eqref{equ:loop-integrands} the form of the loop integrand in terms of one-loop scattering equations,
\begin{equation}
 {\mathfrak{I}}_{\text{NS-NS}}=\int_{\mathfrak{M}_{0,n+2}}\hspace{-20pt}\d\mu_{1,n}^{(\text{nod})}\,\cI^{(1)}_{\text{NS-NS}} \,, \qquad \qquad
\cI^{(1)}_{\text{NS-NS}} =\cI^{(1)}_{\text{NS}}\,\widetilde\cI^{(1)}_{\text{NS}} \,.
\end{equation}
While we did not write it in this way previously, we can use the decomposition 
\begin{equation}
\cI^{(1)}_{\text{NS}} = \frac{1}{(\sigma_{+-})^2}  \left( (D-2) \ \text{Pf}(M_3) + {\text{Pf}}_{\sqrt{q}}(M_3) \right)\,,
\end{equation}
where the definition of ${\text{Pf}}_{\sqrt{q}}(M_3)$, which is not important for our current purpose, can be found in \cite{Geyer:2015bja} or \cite{Geyer:2015jch}. What matters to us now is that the first contribution on the right-hand side is the coefficient of $D-2$, and the other contribution is the rest. Then we can write
\begin{align}
\cI^{(1)}_{\text{NS-NS}} & = \cI^{(1)}_{\text{NS}}\,\widetilde\cI^{(1)}_{\text{NS}} \nonumber \\
& = \frac{1}{(\sigma_{+-})^4} \left( (D-2)^2 \ \text{Pf}(M_3)\ \text{Pf}(\tilde M_3) + (D-2) \, \left[\text{Pf}(M_3)\ {\text{Pf}}_{\sqrt{q}}(\tilde M_3)+{\text{Pf}}_{\sqrt{q}}(M_3)\ \text{Pf}(\tilde M_3)\right] \right.\nonumber \\
& \left. \qquad\qquad\qquad + \text{Pf}_{\sqrt{q}}(M_3)\ {\text{Pf}}_{\sqrt{q}}(\tilde M_3) \right)
\,.
\end{align}
We mentioned in \eqref{equ:loop-integrand-pure} that, for pure gravity in $D=4$, we need to subtract the dilaton and the axion:
\begin{align}
\cI^{(1)}_{\text{4D-pure-grav}} & = \cI^{(1)}_{\text{NS}}\,\widetilde\cI^{(1)}_{\text{NS}} - 2\ \cI^{(1)}_{\text{scal}}\,\tilde\cI^{(1)}_{\text{scal}}  \nonumber \\
& = \frac{1}{(\sigma_{+-})^4} \left( 2 \ \text{Pf}(M_3)\ \text{Pf}(\tilde M_3) + 2\, \left[\text{Pf}(M_3)\ {\text{Pf}}_{\sqrt{q}}(\tilde M_3)+{\text{Pf}}_{\sqrt{q}}(M_3)\ \text{Pf}(\tilde M_3)\right] \right.\nonumber \\
& \left. \qquad\qquad\qquad + \text{Pf}_{\sqrt{q}}(M_3)\ {\text{Pf}}_{\sqrt{q}}(\tilde M_3) \right)
\,.
\end{align}
Since our numerators for NS-NS gravity follow from the expression for $\cI^{(1)}_{\text{NS-NS}}$, it is clear how to modify them in order to subtract the unwanted states: we should modify the coefficient of $(D-2)^2$,
\begin{equation}
N_{\text{4D-pure-grav}} =  N(\{\epsilon_i,k_i\};\ell) \ N(\{\tilde\epsilon_i,k_i\};\ell) \Big|_{\ (D-2)^2 \ \to \ (D-2)^2-2} \quad \text{and set}\quad D=4\,.
\end{equation}
Since graviton external states are in general non-factorisable (i.e.~not of the form $\varepsilon_{\mu\nu}=\epsilon_\mu \epsilon_\nu$), an appropriate symmetric linear combination of numerators for factorisable states $\epsilon_\mu \tilde\epsilon_\nu$ is required. This is not necessary in a four-dimensional helicity basis, where the two helicity states of the graviton have the factorisable form $\varepsilon_{\mu\nu}^\pm=\epsilon_\mu^\pm \epsilon_\nu^\pm$.

The pure gravity numerators obtained in this way also satisfy the previous tests of gauge invariance, independence with respect to reference ordering, and correct higher-point all-plus and one-minus helicity amplitudes.

\section{Discussion} \label{sec:discussion}

We have presented explicit formulae for the one-loop integrands in non-supersymmetric Yang-Mills theory and gravity. The type of integrand representation employed, which differs from a Feynman-like representation with quadratic propagators, has allowed us to extend tree-level results to loop level, and also to check our expressions in a straightforward manner.

These results exhibit the potential of ambitwistor strings as a tool in quantum field theory, including the study of non-supersymmetric theories. One major question that we discussed in detail is the BCJ colour-kinematics duality in gauge theory and the associated double copy to gravity. We found that the one-loop results for supersymmetric theories \cite{He:2016mzd,He:2017spx} extend to the non-supersymmetric case, if one takes into account the fact that the loop integrand is not uniquely defined due to terms that vanish upon loop integration. These results indicate that the loop-level BCJ conjecture \cite{Bern:2010ue}, formulated for a Feynman-type representation of the integrand, is more restrictive than its analogue conjecture for the new type of integrand representation. This is a significant finding in view of the obstacles that have been found when exploring the colour-kinematics duality at loop level, as discussed in the Introduction.

An important open question is how to perform the loop integration efficiently in the representation of the integrand that we use. There are three obvious approaches. The first is to explore the direct loop integration, based on the $i\epsilon$ prescription proposed in Ref.~\cite{Baadsgaard:2015twa}. The second is to adapt our algorithm to produce a Feynman-type representation of the integrand, for which integration techniques have been developed over decades. The third is to extract information from our formulae in a manner that is suitable for modern unitarity techniques. In our view, all three approaches are worth exploring, and indeed each may inform the others.

There are several other lines of work that suggest themselves.  One is the application of our results to infrared physics in gauge theory and gravity, where the scattering equations and ambitwistor strings have already proven to be fruitful \cite{Schwab:2014xua,*Geyer:2014lca,*Adamo:2014yya,*Adamo:2015fwa,*Lipstein:2015rxa} for describing the new soft theorems discovered in \cite{Cachazo:2014fwa,*Casali:2014xpa}. Another line of work is the study of the four-dimensional ambitwistor strings \cite{Geyer:2014fka}, whose loop-level development has been initiated in \cite{Farrow:2017eol} but is still largely an open problem. It would be interesting to find explicit integrand formulae that can in principle be integrated, along the lines of our work.

The most obvious open question is the extension of the entire formalism to higher loops. The first two-loop results were reported in \cite{Geyer:2016wjx} and they will soon be further developed \cite{wip}.

\section*{Acknowledgements}
We would like to thank Simon Caron-Huot, Bo Feng, Song He, Henrik Johansson and in particular Oliver Schlotterer for inspiring discussions and
valuable comments. This research was supported in part by the National Science Foundation under Grant No. NSF PHY-1125915, as well as the Munich Institute for Astro- and Particle Physics (MIAPP) of the DFG cluster of excellence ``Origin and Structure of the Universe''. We are grateful to the KITP Santa Barbara and the MIAPP Munich as well as the organizers of the workshops ``Scattering Amplitudes and Beyond'' (KITP) and ``Mathematics and Physics of Scattering Amplitudes'' (MIAPP) for providing hospitality, support and a stimulating atmosphere. YG gratefully acknowledges support from the National Science Foundation, the W.M. Keck Foundation Fund and the Roger Dashen Membership. RM is supported by a Royal Society University Research Fellowship.



\appendix
\section{Master numerators at four points}\label{sec:numexamples}
For all of this section, we are assuming the reference ordering
$$
\text{RO}= (+1234-) \,.
$$
Following the algorithm given above, two of the master numerators are
\begin{align}
\label{eq:N12341234}
 N(1234) &= (D-2)\, \epsilon_1\cdot\ell\, \epsilon_2\cdot(\ell+k_1) \,\epsilon_3\cdot (\ell-k_4)\,\epsilon_4\cdot \ell\\
  &\qquad +\Big(\tr(12)\epsilon_3\cdot (\ell-k_4)\,\epsilon_4\cdot \ell+\tr(13)\epsilon_2\cdot(\ell+k_1)\,\epsilon_4\cdot\ell +\tr(14) \epsilon_2\cdot(\ell+k_1)\,\epsilon_3\cdot (\ell-k_4)  \nonumber\\
  &\qquad\qquad +tr(23)\epsilon_1\cdot\ell\,\epsilon_4\cdot\ell +\tr(24)\epsilon_1\cdot\ell\,\epsilon_3\cdot (\ell-k_4)+tr(34)\epsilon_1\cdot\ell\,\epsilon_2\cdot(\ell+k_1)\Big) \nonumber\\
  &\qquad - \Big(\tr(123)\epsilon_4\cdot\ell+\tr(124)\epsilon_3\cdot (\ell-k_4)+\tr(134)\epsilon_2\cdot(\ell+k_1) +\tr(234)\epsilon_1\cdot\ell\Big)\nonumber\\
  &\qquad +\tr(1234)\,, \nonumber
\end{align}

\begin{align}
 N(1243) &= (D-2)\,\epsilon_1\cdot\ell\, \epsilon_2\cdot(\ell+k_1) \,\Big(\epsilon_4\cdot (\ell-k_3)\,\epsilon_3\cdot \ell - \epsilon_3\cdot\epsilon_4\,k_4\cdot(\ell-k_3)\Big)\\
  &\qquad +\Big(\tr(12)\big(\epsilon_4\cdot(\ell-k_3)\,\epsilon_3\cdot\ell  - \epsilon_3\cdot\epsilon_4\,k_4\cdot(\ell-k_3)\big) +\tr(13)\epsilon_2\cdot(\ell+k_1)\,\epsilon_4\cdot(\ell-k_3)   \nonumber\\
  &\qquad\qquad +\tr(14) \epsilon_2\cdot(\ell+k_1)\,\epsilon_3\cdot \ell+tr(23)\epsilon_1\cdot\ell\,\epsilon_4\cdot(\ell-k_3) \nonumber\\
   &\qquad\qquad+\tr(24)\epsilon_1\cdot\ell\,\epsilon_3\cdot \ell+tr(34)\epsilon_1\cdot\ell\,\epsilon_2\cdot(\ell+k_1)\Big) \nonumber\\
  &\qquad - \Big(\tr(123)\epsilon_4\cdot(\ell-k_3)+\tr(124)\epsilon_3\cdot \ell+\tr(134)\epsilon_2\cdot(\ell+k_1) +\tr(234)\epsilon_1\cdot\ell\Big) \nonumber\\
  &\qquad +\tr(1243)\,. \nonumber
\end{align}
As discussed in \cref{sec:examples}, the term proportional to $(D-2)$ in $N(1243)$ for example is calculated as the sum over the decompositions $\{\{1\},\{2\},\{3\},\{4\}\}$ and $\{\{1\},\{2\},\{4,3\}\}$,
\begin{equation}
 \epsilon_1\cdot\ell\, \epsilon_2\cdot(\ell+k_1) \,\Big(\epsilon_4\cdot (\ell-k_3)\,\epsilon_3\cdot \ell+\epsilon_3\cdot F_4\cdot(\ell-k_3) \Big)\,.
\end{equation}
Following the remark in \cref{sec:examples}, it is much more convenient to calculate the difference $\Delta=N_{\text{CO}}(\text{CO})-N_{\text{RO}}(\text{CO})$ of the numerators with the reference ordering RO from the canonical reference ordering RO = CO. We detail below the master numerators for the colour orderings (1234) through (1432) in this notation:
\begin{align}
 \Delta(1234) &= 0\,,\\
 \Delta(1243) &= (D-2) \,\epsilon_1\cdot\ell\, \epsilon_2\cdot(\ell+k_1)\,\epsilon_3\cdot\epsilon_4\,k_4\cdot(\ell-k_3) + \tr(12)\epsilon_3\cdot\epsilon_4\,k_4\cdot(\ell-k_3)\,,\\
 \Delta(1342) &= (D-2)\epsilon_1\cdot\ell\,\Big(\epsilon_3\cdot(\ell+k_1)\,\epsilon_2\cdot\epsilon_4 \,k_4\cdot(\ell-k_2)\\
   &\qquad\qquad +k_3\cdot(\ell+k_1)\,\big(\epsilon_2\cdot\epsilon_3 \,\epsilon_4\cdot(\ell-k_2) +\epsilon_3\cdot\epsilon_4 \epsilon_2\cdot k_4 - \epsilon_2\cdot\epsilon_4\, k_4\cdot k_3\big)\Big) \nonumber\\
   &\qquad+\tr(13)\epsilon_2\cdot\epsilon_4\,k_4\cdot(\ell-k_2) +\tr(14)\epsilon_2\cdot \epsilon_3\, k_3\cdot(\ell+k_1) \,,\nonumber\\
 \Delta(1324) &= (D-2) \,\epsilon_1\cdot\ell\, \epsilon_4\cdot \ell\,\epsilon_2\cdot\epsilon_3\,k_3\cdot(\ell+k_1) + \tr(14)\epsilon_2\cdot\epsilon_3\,k_3\cdot(\ell+k_1)\,, \\
 \Delta(1423) &= (D-2)\epsilon_1\cdot\ell\, k_4 \cdot(\ell+k_1)\,\Big(\epsilon_3\cdot \ell\,\epsilon_2\cdot\epsilon_4+\epsilon_2\cdot(\ell+k_1)\,\epsilon_3\cdot\epsilon_4\Big) \\
   &\qquad +\tr(13)k_4 \cdot(\ell+k_1)\,\epsilon_2\cdot\epsilon_4+\tr(12) k_4 \cdot(\ell+k_1)\,\epsilon_3\cdot\epsilon_4\,,\nonumber\\
 \Delta(1432) &= (D-2)\,\epsilon_1\cdot\ell\,\Big(\epsilon_4\cdot(\ell+k_1)\,\epsilon_2\cdot\epsilon_3\,k_3\cdot(\ell-k_2)\\
   &\qquad\qquad +k_4\cdot(\ell+k_1)\,\big(\epsilon_3\cdot\epsilon_4\,\epsilon_2\cdot(\ell-k_4) + \epsilon_2\cdot\epsilon_4\, \epsilon_3\cdot (\ell-k_2) - \epsilon_2\cdot\epsilon_3 \epsilon_4\cdot k_3\big)\Big)\nonumber\\
   &\qquad +\Big(\tr(12)\,\epsilon_3\cdot\epsilon_4 +\tr(13)\,\epsilon_2\cdot\epsilon_4 \Big)\,k_4\cdot(\ell+k_1) + \tr(14)\,\epsilon_2\cdot\epsilon_3 \,k_3\cdot(\ell-k_2)\,.\nonumber
\end{align}


\section{Master numerators independent of the reference ordering}\label{sec:mastersym}
There is a simple way to obtain BCJ master numerators independent of the reference ordering by just summing over all choices:
\begin{equation}
 N_{\text{sym}}(\text{CO}) \equiv \frac{1}{n!}\sum_{\text{RO}}N_{\text{RO}}(\text{CO})\,.
\end{equation}
For four external particles, the resulting numerators can be expressed (relatively) compactly as
\begin{equation}
 N_{\text{sym}}(1234) = N_{(1234)}(1234)+(D-2)\Delta_{\text{sym}}^{(D-2)}+\Delta_{\text{sym}}^{\text{tr}}\,,
\end{equation}
where $N_{(1234)}(1234)$ was given in \eqref{eq:N12341234}, and the contributions from contractions among polarisation vectors are 
\begin{subequations}
\begin{align}
 \Delta_{\text{sym}}^{(D-2)} &= \frac{1}{12}\Bigg[3\, \epsilon_4\cdot p_4 \,\,\mathcal{U}_1\big(3\,|\,2,1\,|\,\ell\big) +\epsilon_4\cdot \ell \,\,\mathcal{U}_1\big(3\,|\,2,1\,|\,p_1\big) \,\, +(1243)+(1342)+(2341)\Bigg]\\
 &\qquad\qquad +\,\,\frac{1}{8}\,\Bigg[  \mathcal{U}_2\big(\,2,1\,|\,\ell\,\big\Arrowvert\,4,3\,|\,p_3\big) + \mathcal{U}_2\big(\,2,1\,|\,p_1\,\big\Arrowvert\,4,3\,|\,\ell\big) \,+ (1423)+(1324)\Bigg] \nonumber\\
 &\qquad\qquad -\frac{1}{24}\Bigg[\epsilon_2\cdot \epsilon_1 \,k_1\cdot p_1 \,\epsilon_3\cdot\ell \,\Big(2\,\epsilon_4\cdot p_4+ 2\,\epsilon_4\cdot \ell+\epsilon_4\cdot k_3\Big)\nonumber\\
 &\qquad\qquad\qquad + \epsilon_2\cdot \epsilon_1\,k_1\cdot \ell \,\epsilon_3\cdot p_3 \,\Big(6\,\epsilon_4\cdot p_4+2\,\epsilon_4\cdot \ell-3\,\epsilon_4\cdot k_3\,\Big)\nonumber\\
 &\qquad\qquad\qquad + (1324)+(1423)+(2314)+(2413)+(3412)\Bigg]\nonumber\\
 &\qquad\qquad +\frac{1}{4}\,\mathcal{U}_1\big(4\,|\,3,2,1\,|\,\ell\big) + \frac{1}{12}\epsilon_4\cdot \epsilon_3\,k_3\cdot k_2\,\epsilon_1\cdot\ell\,\epsilon_2\cdot k_1\,,\nonumber\\
 \Delta_{\text{sym}}^{\text{tr}} &= -\frac{1}{2}\Bigg[\tr\big(12\big)\,\,\epsilon_4\cdot\epsilon_3\,\, k_3\cdot p_3 + (1324)+(1423)+(2314)+(2413)+(3412)\Bigg]\,.
\end{align}
\end{subequations}
Above, we used the convention
\begin{equation}
 p_1 = \ell\,,\qquad\qquad p_2 = \ell+k_1\,,\qquad\qquad p_3 = \ell-k_4\,\qquad\qquad \text{and }\,\, p_4 = \ell\,,
\end{equation}
and the short-hand notation
\begin{subequations} 
\begin{align}
 & \mathcal{U}_1\big(a_n\,|\,a_{n-1},\,...,\,a_1\,|\,p\big) = \epsilon_{a_n}\cdot F_{a_{n-1}}\cdot...\cdot F_{a_1}\cdot p - \epsilon_{a_n}\cdot k_{a_{n-1}}\,\epsilon_{a_{n-1}}\cdot k_{a_{n-2}}\,...\,\epsilon_{a_2}\cdot k_{a_1}\,\epsilon_{a_1}\cdot p\,,\\
 & \mathcal{U}_2\big(\,a_2,a_1\,|\,p\,\big\Arrowvert\,a_4,a_3\,|\,q\big) = \epsilon_{a_2}\cdot F_{a_1}\cdot p\,\,\epsilon_{a_4}\cdot F_{a_3}\cdot q - \epsilon_{a_2}\cdot k_{a_1}\,\epsilon_{a_1}\cdot p\,\epsilon_{a_4}\cdot k_{a_3}\,\epsilon_{a_3}\cdot q\,.
\end{align}
\end{subequations}
Moreover, we denote the permutations summed over by e.g.
\begin{equation}
 (2413)=\begin{pmatrix}1&2&3&4\\2&4&1&3\end{pmatrix}\equiv (\,1\rightarrow 2,\,2\rightarrow 4,\,3\rightarrow 1,\,4\rightarrow 3\,)\,.
\end{equation}

\bibliography{twistor-bib}
\bibliographystyle{JHEP_mod}

\end{document}